# Computational Design of Crescent Shaped Promising Non-Fullerene Acceptors with 2,3-quinoxaline,1,4-dihydro Core and Different Electron-withdrawing Terminal Units for Photovoltaic Applications

Labanya Bhattacharya,[1] Alex Brown[2], Sagar Sharma,[3] and Sridhar Sahu[1*]

[1]*Computational Materials Research Lab, Department of Physics, Indian Institute of Technology (Indian School of Mines), Dhanbad, Jharkhand-826004, India*

[2]*Department of Chemistry, University of Alberta, Edmonton, Alberta, T6G 2G2, Canada*

[3]*Department of Chemistry, S. B. Deorah College, Bora Service, Ulubari, Guwahati-781007, Assam, India*

E-mail: *sridharsahu@iitism.ac.in

## Abstract

This study aims to design a series of non-fullerene acceptors (NFAs) for photovoltaic applications having 2,3-quinoxalinedione,1,4-dihydro fused thiophene derivative as the core unit and 1,1-dicyanomethylene-3-indanone (IC) derivatives and different $\pi$-conjugated molecules other than IC as terminal acceptor units. All the investigated NFAs are found air-stable as the computed highest occupied molecular orbitals (HOMOs) are below the air oxidation threshold (ca. -5.27 eV vs. saturated calomel electrode). The studied NFAs can act as potential non-fullerene acceptor candidates as they are found to have sufficient open-circuit voltage ($V_{oc}$) and fill factor (FF) ranging from 0.62-1.41 eV and 83%-91%, respectively. From the anisotropic mobility analysis, it is noticed that the studied NFAs except dicyano-rhodanine terminal unit containing NFA, exhibit better electron mobility than the hole mobility, and therefore, they can be more promising electron transporting acceptor materials in the active layer of an organic photovoltaic cell. From the optical absorption analysis, it is noted that all the designed NFAs have the maximum absorption spectra ranging from 597 nm-730 nm, which lies in the visible region and near infra-red (IR) region of the solar spectrum. The computed light-harvesting efficiencies for the PM6 (thiophene derivative donor selected in our study): NFA blends are found to lie in the range of 0.9589-0.9957, which indicates efficient light-harvesting by the PM6:NFA blends during photovoltaic device operation.

# INTRODUCTION

In recent years, acceptors have been at the forefront of organic-based photovoltaic research and development.[1,2,3,4] So far, two types of acceptors have been extensively studied and they are fullerene acceptors (FAs) and non-fullerene acceptors (NFAs). Over the last few years, fullerene and its derivatives such as 6,6-phenyl-C61-butyric acid methyl ester ($PC_{61}BM$), 6,6-phenyl-C71-butyric acid methyl ester ($PC_{71}BM$), and indene-$C_{60}$ bisadduct (ICBA) played significant roles in boosting the organic photovoltaic performance as they all have a high electron affinity, unique spherical geometry, and good isotropic charge transport ability.[5,6,7,8,9] However, their difficult and expensive synthesis, limited structural tunability, weak absorption spectra, and thermal instability become major bottlenecks for improving device performance.[10,11,12] Alternatively, non-fullerene acceptors, namely 3,9-bis(2-methylene-(3-(1,1-dicyanomethylene)-indanone))-5,5,11,11-tetrakis(4-hexylphenyl)-dithieno[2,3-d:2´,3´-d´]-s-indaceno[1,2-b:5,6-b´]dithiophene (ITIC), 2,2´-((2Z,2´Z)-((4,4,9,9-tetrahexyl-4,9-dihydro-s-indaceno[1,2-b:5,6-b´]dithiophene-2,7-diyl)bis(methanylylidene))bis(3-oxo-2,3-dihydro-1H-indene-2,1-diylidene))dimalononitrile (IDIC), and 2,2´-((2Z,2´Z)-((12,13-bis(2-ethylhexyl)-3,9-diundecyl-12,13-dihydro-[1,2,5]thiadiazolo[3,4-e]thieno[2´´,3´´:4´,5´]thieno[2´,3´:4,5]pyrrolo[3,2-g]thieno[2´,3´:4,5]thieno[3,2-b]indole-2,10-diyl)bis(methanylylidene))bis(5,6-difluoro-3-oxo-2,3-dihydro-1H-indene-2,1-diylidene))dimalononitrile (Y6) have several advantages over fullerene acceptors such as easy synthesis, tunable energy levels via simple structural modifications, extended optical absorption profiles and favorable charge transfer at low eneregtic driving force while maintaining a good open-circuit voltage to short-circuit current density tradeoff.[13,14,15] Non-fullerene acceptors typically contain different donor (D) and acceptor (A) blocks as core and terminal units, along with (possibly) $\pi$-bridging units with arrangements such as A-D-A, A-$\pi$-D-$\pi$-A, $A_2$-$DA_1D$-$A_2$. Different design strategies of core, terminal and bridging units, as well as their arrangements within the molecule, have led to promising NFAs for photovoltaic applications.[13,16,17] In particular, structural tuning of core and terminal (or end) groups has been reported to be a promising strategy for building potential NFAs. For example, indacenodithiophene (IDT) is one of the most widely used cores in the NFA backbone. When the central benzene ring of the IDT core was substituted by naphthalene, the weaker electron-donating ability led to a blueshifted optical absorption, and the power conversion efficiency (PCE) of the devices having PBDB-T (donor):NFA reached up to 8-9%.[18,19] In the IDT core, if the central benzene is replaced by an electron-rich thieno[3,2-*b*]thiophene (TT) unit, stronger intramolecular charge transfer (ICT) has been observed due to efficient $\pi$-electron delocalization.[20,21] In addition, the stronger ICT broadened the absorption into the near-infrared (NIR) region, strong intermolecular interactions led to ordered molecular stacking, and the PCEs reached over 10% when matched with the PTB7-Th donor.[20,21] In 2015, Lin *et al.* reported a NFA, namely ITIC, which achieved a PCE of up to 6.8% when blended with the PTB7-Th donor.[22] Electron withdrawing fluorine atom substitution to the terminal 1,1-dicyanomethylene-3-indanone (IC) unit of ITIC leads to low-lying frontier molecular energy levels, a red-shifted absorption profile, and improved device efficiency.[23] Inspired by ITIC in 2017, Zou's group designed a new NFA named BZIC (dithieno[3,2-*b*]-pyrrolobenzotriazole-1,1-dicyanomethylene-3-indanone), having a nitrogen-atom-substituted core with a red-shifted optical absorption profile.[24] Combining the advantages of ITIC and BZIC, Yang's group designed the Y-series NFAs such as Y1, Y3, and Y6 (Y is an abbreviation for Yang).[25] The conjugation length of the nitrogen-atom-substituted core in BZIC is increased by thiophene ring incorporation in Y1, and hence ICT is enhanced, and a red-shifted absorption spectrum was measured. Y3 was designed via terminal group modulation of Y1. Y1 contains 1,1-dicyanomethylene-3-indanone (IC), in which they substituted electron withdrawing fluorine (F) atoms, and the electronegativity of fluorine atoms in Y3 enhances ICT, broadens the optical absorption profile, and improves the molecular packing. In Y6, the central unit of the D-A-D core in BZIC, Y1, and Y3, i.e., benzotriazole (BTZ), is replaced by the stronger electron withdrawing benzothiadiazole (BT). In comparison to Y1 and Y3, Y6 showed superior morphology, high charge carrier mobility, low-lying highest occupied molecular orbital (HOMO) and lowest unoccupied molecular orbital (LUMO) levels, narrower bandgap, and a red-shifted absorption spectrum. The core and terminal modulation design strategy of crescent shaped Y series NFAs provides a PCE gain of 13.4% (Y1), 14.8% (Y3) to 15.7% (Y6), as well as an increase in fill factor from 69.1% (Y1), 71.2% (Y3) to 74.8% (Y6).[25] Recently, the highest efficiency for single junction organic solar cells is reported as 18.22%, where the active layer is composed of D18 polymer donor and Y6 non-fullerene acceptor.[26] Khan *et al.* computationally designed new Y-series near-infrared sensitive NFAs based on reported Y21 acceptor (asymmetric core D-A-D unit) via terminal group modulation.[27] They found that terminal unit variation reduced the frontier molecular orbital energy levels, reorganization energy, exciton binding energy, as well as improved the absorption maximum and open-circuit voltage. Recently, Li and his group theoretically designed a series of six NFAs (Y6-COH, Y6-COOH, Y6-CN, Y6-SO$_2$H, Y6-CF$_3$, and Y6-NO$_2$). They replaced the electronegative fluorine (F) atoms on the terminal fluorinated IC unit of Y6 with different electron-withdrawing end groups (-COH, -COOH, -CN, -SO$_2$H, -CF$_3$, -NO$_2$).[28] They observed that among the end-capped engineered NFAs, Y6-NO$_2$ has the lowest-lying frontier molecular orbitals and the largest red-shifted absorption profile; moreover, the PM6 (donor)/Y6-NO$_2$ (NFA) complex system showed stronger interfacial interactions and enhanced charge-transfer (CT) characteristics as compared to the PM6/Y6 composite.[28] In 2019, inspired by the excellent Y-series NFAs, Zhou *et al.* introduced a quinoxaline moiety core replacing the benzothiadiazole core of Y6 and designed two Y-series molecules namely, AQx-1 and AQx-2.[29] The quinoxaline moiety is reported to have the quinoid-resonance effect, and the reduced free $\sigma$-bonds in AQx could lower the reorganization energy and improve the electron transport and intermolecular packing. They achieved a device efficiency of 16.64% with significant short-circuit current density and fill factor for the binary organic solar cells containing AQx-2 as NFA and PBDB-TF (PM6) as donor.[29] While the Y-series NFAs have been widely used as active layer components in bulk-heterojunction organic solar cells, and ternary solar cells, there is still considerable room for improvement in non-fullerene acceptors.

In our study, we have modified the recently reported AQx-2 by designing a NFA namely **AQx-2-c** in which quinoxaline core of AQx-2 is replaced by 2,3-quinoxalinedione,1,4-dihydro, which has a weaker electron donating ability than quinoxaline. Further, we performed terminal unit engineering of **AQx-2-c** (terminal unit: Fluorinated IC) with various halogen substituted, $\pi$-extended, and theinyl fused IC groups as well as explored end-capped groups different than IC.[16,30] Thus, we proposed eight new NFAs namely **AQx-2-ct1** (terminal unit: Chlorinated IC), **AQx-2-ct2** (terminal unit: Fluorinated naphtyl fused IC), **AQx-2-ct3** (terminal unit: Chlorinated naphtyl fused IC), **AQx-2-ct4** (terminal unit: $\gamma$-thienyl IC), **AQx-2-ct5** (terminal unit: $\beta$-thienyl IC). **AQx-2-ct6** (terminal unit: Rhodanine), **AQx-2-ct7** (terminal unit: Dicyano-Rhodanine), and **AQx-2-ct8** (terminal unit: Malononitrile). We studied computationally the geometrical, optoelectronic, intra and intermolecular charge transport properties, and overall photovoltaic performance of the designed crescent-shaped NFAs using density functional theory

(DFT) and time-dependent DFT (TD-DFT). In the present study, Poly[(2,6-(4,8-bis(5-(2-ethylhexyl-3-fluoro)thiophen-2-yl)-benzo[1,2-*b*:4,5-*b*´]dithiophene))-alt-(5,5-(1´,3´-di-2-thienyl-5´,7´-bis(2-ethylhexyl)benzo[1´,2´-*c*:4´,5´-*c*´]dithiophene-4,8-dione)] (PM6) is selected as the donor as it has well matched frontier molecular orbitals with respect to the designed NFAs. The absorption properties as well as charge transfer analysis of the PM6 donor and designed NFAs blends or complexes have been investigated.

## COMPUTATIONAL METHODOLOGY

Within the framework of density functional theory, benchmarking a suitable functional to obtain satisfactory agreement with the experimental results available from cyclic voltammetry, and ultraviolet photoelectron spectroscopy measurements has been a difficult task.[28,31,32] In our present study, we optimized the reference NFA, AQx-2 (Figure 1) using different functionals (B3LYP, B3LYP-D3BJ, PBE0, MPW91PW91, M06, and $\omega$B97XD),[33,34,35,36,37,38] coupled with the 6-31G(d,p) and 6-311G(d,p) basis sets to compute the highest occupied molecular orbital (HOMO). The computed HOMO energy (-5.52 eV) of AQx-2 using B3LYP-D3BJ/6-31G(d,p) as listed in Table S1 in the supporting information (SI) was found to be in good agreement with the experimentally reported value (-5.62 eV). The B3LYP functional has been used to optimize the ground state geometries of crescent shaped Y-series NFAs, and the computed HOMO energy is usually found to be in good agreement with experimentally reported results. [28,39,40] From Table S1, the computed HOMO energy is, perhaps not surprisingly, almost the same for B3LYP and B3LYP-D3BJ. Therefore, in our study, the ground state geometries of the designed NFAs (Figure 2) are optimized using the dispersion-corrected B3LYP-D3BJ functional in conjunction with the 6-31G(d,p) basis set in the gas phase. The ground state geometry of the PM6 donor has been determined at the same theoretical level and the highest occupied molecular orbital (HOMO) and lowest unoccupied molecular orbital (LUMO) energies as well as optical band gap ($E_{g,opt}$) are determined by linear extrapolation to $n = \infty$ from the explicit computations on PM6 systems with n=1-4 monomer units (Figure S1, Table S2). The optimized geometries of the studied NFAs and PM6 donor were confirmed as minima by carrying out vibrational frequency analysis, which yielded real harmonic frequencies. For all the studied compounds, the full alkyl chains were retained in calculations.

Further, to choose a reasonable computational level to simulate the optical absorption properties of the investigated NFAs, the maximum absorption wavelength of the reference AQx-2 acceptor (at the B3LYP-D3BJ optimized geometry) was computed using different hybrid (B3LYP-D3BJ, PBE0, M06) and long-range corrected functionals (LC-$\omega$PBE, LC-BLYP, CAM-B3LYP, $\omega$B97XD) in conjunction with the 6-31G(d,p) basis set in Chloroform solvent medium, such that comparison could be made to the reported experimental values of Zhou *et al.* [29] The excitation energies and corresponding oscillator strengths were determined within the framework of TD-DFT alongside the integral equation formalism polarizable continuum (IEF-PCM) model to treat solvation. The simulated maximum absorption wavelength of AQx-2 using the B3LYP-D3BJ/6-31G(d,p) theoretical level was close to the experimental result (Table S3). A similar observation was reported earlier for crescent shaped Y-series NFAs.[27,28] Hence, optical absorption spectra of the structurally similar NFAs investigated here are computed at the B3LYP-D3BJ/6-31G(d,p) theoretical level. Usually, the computed LUMOs at the DFT level differ significantly from experimentally reported values. Therefore, in our study, to predict reliable LUMO energies, we use $E_{LUMO} = E_{HOMO} + E_{g,opt,TD}$, where $E_{g,opt,TD}$ is the optical band gap, evaluated from TD-DFT analysis.[31,32] The reorganization energy, ionization potential, and electron affinity of all the designed NFAs are computed at the B3LYP-D3BJ/6-31G(d,p) level of theory. Crystal structures of the designed NFAs have been predicted employing the most reliable Dreiding force field within the Polymorph Predictor module of the BIOVIA Materials Studio17 based on their optimized gas-phase conformations.[41] All possible crystal structures are selected by energy minimization of crystals with the most common space groups such as *P2₁/C*, *P*1, *P*1, *Pbca*, *P2₁2₁2₁*, *P2₁*, *C2/c*, *Pna2₁*, *Cc*, *Pbcn*, and *C*2.[4,42,43,44,45] Crystal structure prediction method within the Polymorph Predictor module relies on Monte Carlo simulation which is a stochastic process. To assess the completeness of the search within this module, we have repeated the simulations for several times for each initial molecular structure until the same low energy crystal structures are produced. We generated large number of possible polymorphs using Dreiding force field parametrization which is considered to be a reasonable choice for organic molecules.[4,43,44,45] The charge transfer integrals of the designed NFAs are evaluated using the site energy correction method at the PW91/TZP level of theory using the ADF program.[46,47] The PW91/TZP theoretical level has been widely used to evaluate charge transfer integrals of organic molecules.[48,49,50] Further, to model the interface between the PM6 donor and the NFAs, we performed end-to-end acceptor-acceptor (A-A) stacking. In the A-A stacking, the stacking occurs between the electron-withdrawing end groups of the NFAs and the electron withdrawing unit (A) of the polymer PM6 donor. For simplification, only one repeat unit of the polymer donor PM6 was considered in our study, and unsaturated atoms were hydrogenated. The geometry of the bimolecular system (PM6:NFA) is optimized using the computationally efficient extended tight-binding (GFN2-xTB) method.[51] To investigate the intermolecular charge transfer (CT) states of the PM6-NFA complexes, natural transition orbital (NTO) analysis and interfragment charge transfer (IFCT) have been performed using TD-DFT. The NTO and IFCT analysis have been done using the range-separated functional CAM-B3LYP (with the 6-31G(d,p) basis set), as this approach has been widely accepted for calculating the intermolecular CT properties of such donor/acceptor complex systems.[2,52,53] The NTO analysis, transition density matrix (TDM), and charge density difference (CDD) calculations reported in this study were done using Multiwfn 3.8 (dev).[54] Unless otherwise noted, all the DFT and TD-DFT calculations were carried out using the Gaussian 16 program.[55]

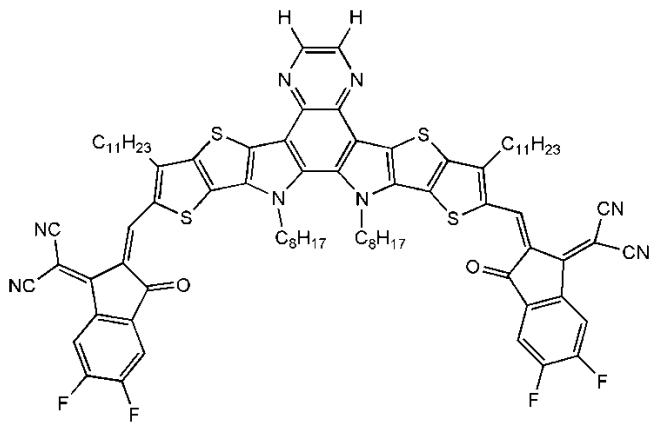

**Figure 1:** Molecular structure of the reference non-fullerene acceptor AQx-2.[29]

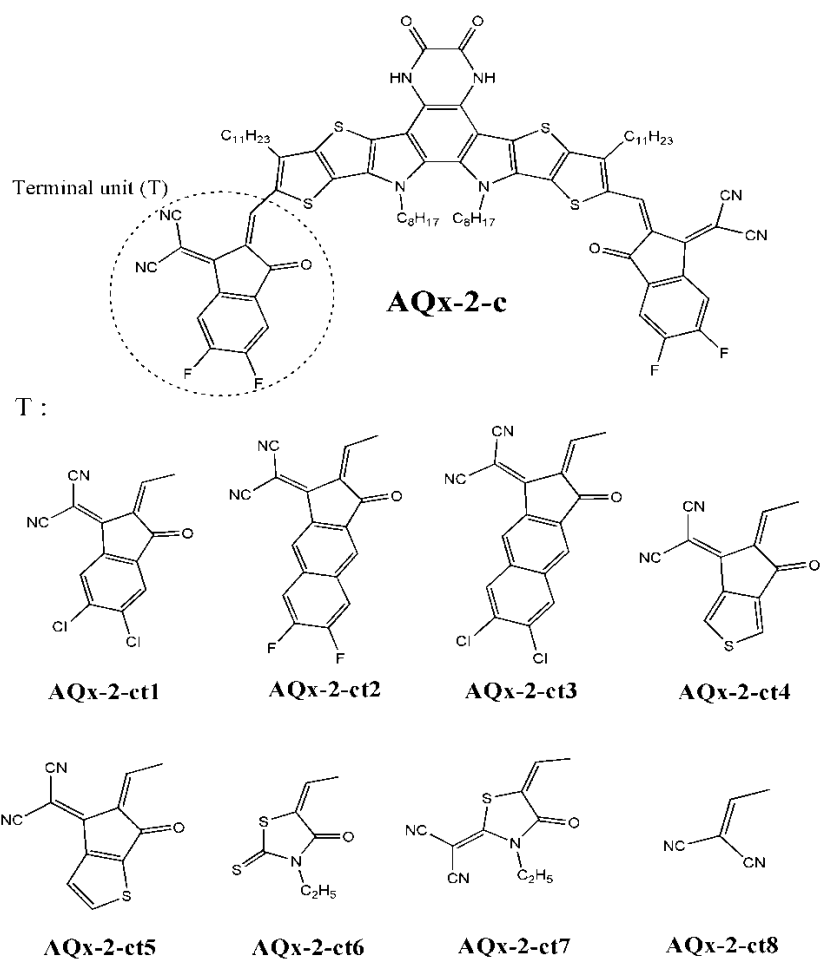

Figure 2: Chemical structures of the designed NFAs.

## RESULTS AND DISCUSSIONS

### Molecular Geometries and Molecular Electrostatic Potential (MEP) of the Designed NFAs.

The chemical structures of the designed NFAs are shown in Figure 2. Important geometrical parameters related to NFA functionality involve the bond lengths ($d_1, d_2, d_3, d_4$) and dihedral angles ($\alpha, \beta, \gamma$) and they have been shown in Figure 3; the optimized parameters are summarized in Table S4. The dihedral angle $\gamma$ (Figure 3) denotes the planarity of the core D-A$_1$-D unit, and $\alpha, \beta$ represents the dihedral angles between the center donor unit and two terminal acceptor units. For all the designed NFAs except **AQx-2-ct8**, $\alpha$ (range: 1.31-1.45°), $\beta$ (range: 1.83-1.96°), and $\gamma$ (range: 1.68-1.94°) values are almost zero indicating the near planarity of the acceptors. It is also observed that all the newly designed NFAs have a modestly more planar core than the reference AQx-2 as they have smaller $\gamma$ values compared to AQx-2 (ca 2° vs 3°). The computed bond lengths range from 1.37 Å to 1.42 Å, that falls between the carbon-carbon single bond length (C-C = 1.54 Å) and the carbon-carbon double bond length (C=C = 1.33 Å). Therefore, all the bonds exhibit significant double bond character, which facilitates the delocalization of π-electrons over the whole molecular backbone and, importantly, this effective π-conjugation increases the charge transport properties.[1]

To realize the variation of electron density over the molecular backbone of the designed NFAs, molecular electrostatic potential (MEP) analysis has been carried out (Figure S2). The MEP plots represent different values of the electrostatic potential in terms of different colors where the negative potential increases on moving from blue to red as blue < green < yellow < orange < red. For **AQx-2-c**, the core $A_1$ unit is more electronegative as it contains electron rich oxygen atom rather than hydrogen in reference NFA AQx-2 (Figure S2). For **AQx-2-c** and other designed NFAs, the negative potential is concentrated mostly on the core unit and the positive potential on the whole conjugated backbone and on the terminal units. Therefore, the separation between the negative and positive regions is more prominent for our designed NFAs than the reference AQx-2 which indicates delocalization of the π-electrons over the molecular backbone.

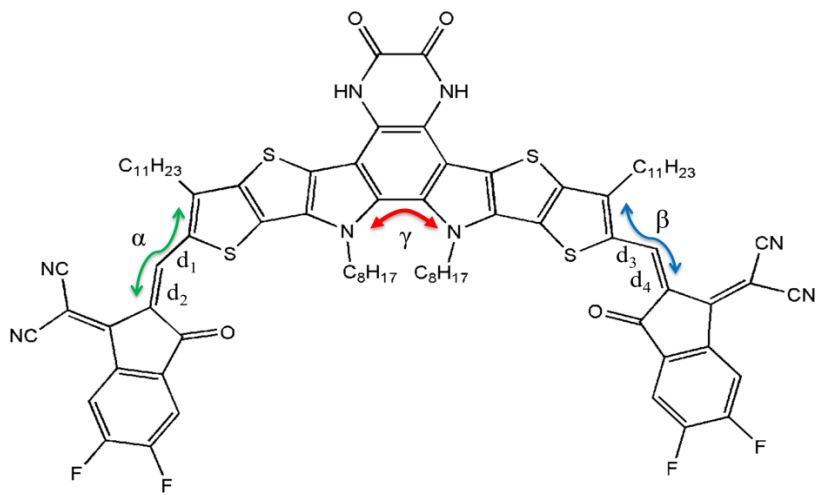

**Figure 3:** Representation of dihedral angles (α, β, γ) and bond lengths ($d_1$, $d_2$, $d_3$, $d_4$) in a NFA (**AQx-2-c**).

### Electronic Properties and Air-stability.

The energies of frontier molecular orbitals, i.e., the HOMO and LUMO energies, are significant factors for estimating the air-stability, optoelectronic, charge transfer, and photovoltaic properties of organic molecules. The HOMO values of the designed NFAs are computed at the B3LYP-D3BJ/6-31G(d,p) level of theory in the gas phase. The LUMO energies are predicted at the same theoretical level in the gas phase using the indirect approach, as discussed in computational methodology section. For organic molecules to be air-stable, the HOMO energy should be below the air oxidation threshold (ca. -5.27 eV or 0.57 V vs. saturated calomel electrode (SCE)).[56] In our present study, the computed HOMO levels of the investigated designed NFAs range from -5.28 to -5.77 eV (Figure 4), and hence, they should possess ambient stability under atmospheric conditions. All the designed NFAs, except **AQx-2-ct6**, possess deeper HOMO energies than that of reference AQx-2, and therefore, they are more air-stable than AQx-2. The chlorine atom has a stronger electron-withdrawing ability than fluorine, and therefore, both **AQx-2-ct1** (1.78 eV) and **AQx-2-ct3** (1.76 eV) show narrow band gaps and downshifted HOMO/LUMO energies compared to their fluorinated counterparts **AQx-2-c** (1.82 eV) and **AQx-2-ct2** (1.79 eV). The NFA with thienyl fused IC terminal units, **AQx2-ct4** (1.82 eV), exhibits a modestly smaller band gap than AQx-2 (1.85 eV). In **AQx-2-ct7**, the electron-withdrawing cyano group in the Rhodanine terminal group leads to low-lying HOMO and LUMO energies and results in improved air-stability as well as a slightly narrowed-band gap acceptor as compared to **AQx-2-ct6**. From Figure 4, the HOMO and the LUMO energies of the designed NFAs are lower-lying than those of the donor PM6 polymer, and hence, they can receive negatively charged electrons from or deliver positively charged holes to the PM6 polymer donor. The contours of the frontier molecular orbitals (FMOs) are shown in Figure S3 and Figure S4, and for all the designed NFAs, the HOMOs are delocalized over the entire conjugated molecular backbone, which suggests free electron availability and potential charge transport capacity. In contrast, the LUMO electron density localizes predominantly on the terminal acceptor units of the designed NFAs, and that indicates transportation of electrons from the core D-$A_1$-D unit to the terminal $A_2$ units. For an n-type air-stable organic molecule, the LUMO energy must approach approximately -4.0 eV to ensure good coupling with the cathode electrode material, such as Gold (Au) or Aluminum (Al).[57] The computed LUMO levels of the designed compounds range from -3.19 eV to -3.98 eV. Except for **AQx-2-ct6**, all the designed NFAs have more n-type character. In general, for organic solar cell devices, Aluminum (workfunction, Φ= -4.3 eV) is selected as the cathode electrode material, and, thus, the charge (hole and electron) injection barriers of the investigated NFAs are computed with respect to the Aluminum (Al) electrode (Table S5). The electron injection barriers of the NFAs except **AQx-2-ct6** are significantly less than the corresponding hole injection barriers indicating n-characteristics of the studied NFAs. The electron injection barriers of the designed NFAs, except **AQx-2-ct5** to **AQx-2-ct8** are less than that for the reference NFA, AQx-2, i.e., 0.32-0.49 eV < 0.63 eV; and thus **AQx-2-c** and **AQx-2-ct1** to **AQx-2-ct4** will couple well with an Al electrode. As a result, those NFAs may favor better electron injection than AQx-2.

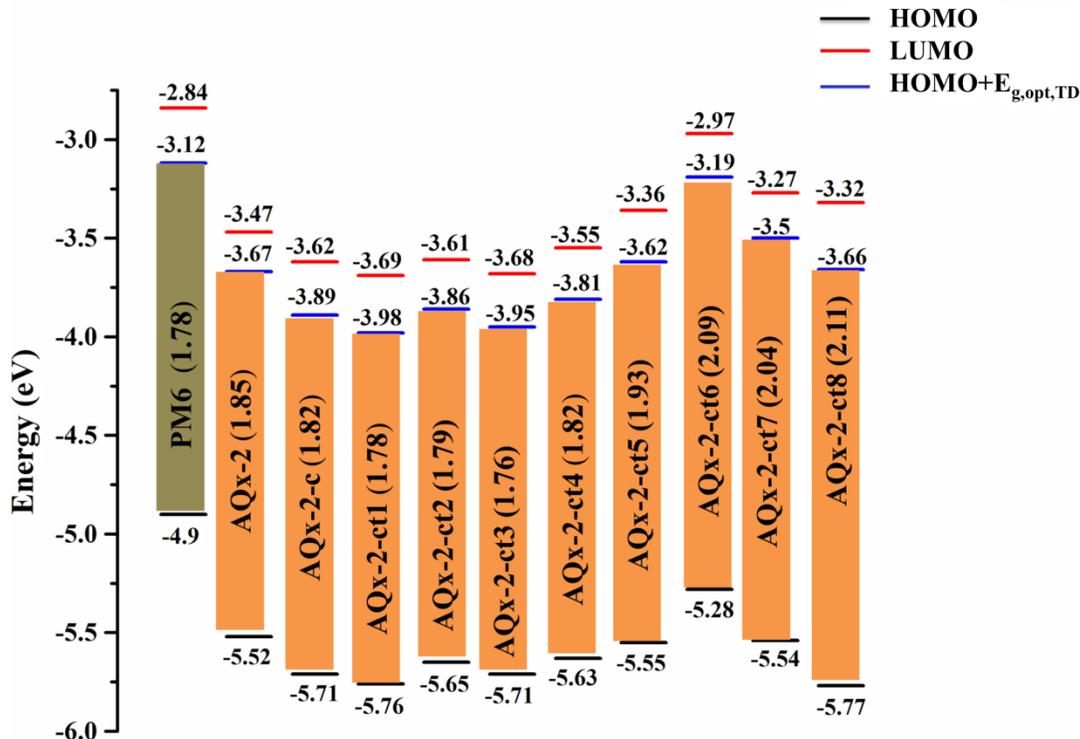

**Figure 4:** HOMO and LUMO (DFT and TDDFT both with B3LYP-D3BJ/6-31G(d,p)) energies of the designed NFAs and PM6 donor in the gas phase.

**Table 1:** Computed open-circuit voltage ($V_{oc}$), Fill Factor (FF), Energetic Driving Force ($L_D$-$L_A$) Between PM6 Donor and Designed NFAs at B3LYP-D3BJ/6-31G(d,p) Level of Theory

| Comp. | $V_{oc}$ (V) | FF (%) | $L_D$-$L_A$ (eV) |
|---|---|---|---|
| AQx-2 (Ref.) | 0.93 | 87 | 0.55 |
| **AQx-2-c** | 0.71 | 85 | 0.77 |
| **AQx-2-ct1** | 0.62 | 83 | 0.86 |
| **AQx-2-ct2** | 0.74 | 85 | 0.74 |
| **AQx-2-ct3** | 0.65 | 84 | 0.83 |
| **AQx-2-ct4** | 0.79 | 86 | 0.69 |
| **AQx-2-ct5** | 0.98 | 88 | 0.50 |
| **AQx-2-ct6** | 1.41 | 91 | 0.07 |
| **AQx-2-ct7** | 1.10 | 89 | 0.38 |
| **AQx-2-ct8** | 0.94 | 88 | 0.54 |

## Open-circuit Voltage ($V_{oc}$), Fill Factor, and Energetic Driving Force.

The key photovoltaic parameters of an organic solar cell, including open-circuit voltage ($V_{oc}$), fill factor (FF), short-circuit current density ($J_{sc}$), and driving force energy ($L_D$-$L_A$) are highly correlated with the HOMO and LUMO values of the donors and acceptors. The open circuit voltage $V_{oc}$ for the studied NFAs are estimated from the empirical equation proposed by Scharber et al.,[58]

$$V_{oc} = (\tfrac{1}{e})(|E_{HOMO}(D)| - |E_{LUMO}(A)|) - 0.3\,V \quad (1)$$

Here e is the elementary charge, $E_{HOMO}(D)$ is the HOMO level of the donor (D) and $E_{LUMO}(A)$ is the LUMO level of the acceptor (A). The empirical factor, 0.3 V represents losses in transport to the electrodes.[58] It is evident that a deeper HOMO level of the donor combined with a shallower LUMO level of the acceptor leads to high $V_{oc}$. For example, **AQx-2-c** has an upshifted LUMO energy compared to its chlorinated counterpart **AQx-2-ct1**, and, hence, **AQx-2-c** has a higher $V_{oc}$ (0.71 eV) than **AQx-2-ct1** (0.62 V). Similarly, the LUMO energy is found to be higher for **AQx2-ct5** having β-thienyl IC terminal group than **AQx-ct4** (terminal unit: γ-thienyl IC), and that leads to higher $V_{oc}$ for **AQx2-ct5** (0.98 V) as compared to **AQx-2-ct4** (0.79 V). It is reported that during the photo-oxidation process in blend films, NFAs having low-lying LUMO levels could suppress the charge transfer between NFAs and oxygen, and that leads to quenching photo-oxidation for both donors and acceptors in blend films, which is important for the stability of NFAs.[28] In this present study, among all the designed NFAs, **AQx-2-c**, **AQx-2-ct1**, **AQx-2-ct2**, **AQx-2-ct3**, **AQx-2-ct4**, **AQx-2-ct5**, **AQx-2-ct8** have lower-lying LUMO energies, and therefore, they have excellent photo-oxidation stability but relatively smaller $V_{oc}$. Nevertheless, the computed $V_{oc}$s of the designed NFAs as listed in Table 1 range from 0.62-1.41 eV. The computed $V_{oc}$ for the reference NFA AQx-2 is 0.93 V, which is slightly overestimated relative to the reported experimental value (0.86 V).[29] This slight deviation between computed and measured values is reasonable because experimentally $V_{oc}$ depends on energy levels as well as temperature, light source, charge-carrier recombination, light intensity, and morphology of the photovoltaic device.[1]

The fill factor (FF) is one of the essential parameters that affect the power conversion efficiency of the organic photovoltaic device. However, series resistance ($R_s$) and shunt resistance ($R_{sh}$) are the two critical factors that influence the FF.[59] For organic materials, theoretical prediction of the $R_s$ and $R_{sh}$ is still a challenging task. Therefore, in ideal condition, FF can be expressed as [60,61],

$$FF = \frac{\gamma_{oc} - \ln(\gamma_{oc} + 0.72)}{\gamma_{oc} + 1} \quad (2)$$

Here, $\gamma_{oc}$ is defined as dimensionless voltage:

$$\gamma_{oc} = \frac{eV_{oc}}{k_B T} \quad (3)$$

In this equation, e, $V_{oc}$, $k_B$, and T denote elementary charge, open-circuit voltage, Boltzmann constant and temperature (300 K), respectively. The accuracy of the equation holds for $\gamma_{oc}$>10.[60,61] The calculated $\gamma_{oc}$ values for the studied NFAs are found to be greater than 10. The computed fill factor values are listed in Table 1. The FF is linearly correlated with the $V_{oc}$, and therefore, the NFAs having upshifted LUMO levels have comparatively higher FF.[59,60,61] The computed fill factor for AQx-2 is 87%, which is slightly overestimated compared to the reported experimental reported value (76%).[29] In practical operating conditions, various loss mechanisms also take place, which are neglected during the theoretical estimation of the FF values.[61] Therefore, the computed FF values are usually found to be overestimated, but the predicted trends are expected to be the similar.

In our study, the calculated LUMO-LUMO offsets or the energetic driving forces ($L_D$-$L_A$), i.e., the LUMO energy differences between the donor and acceptor are listed in Table 1. The NFAs having upshifted LUMO levels possess low energetic driving forces. Therefore, they may facilitate the fast exciton dissociation process at the donor/acceptor interface as compared to others with higher LUMO-LUMO offsets.

## Optical Absorption Properties.

The short-circuit current density ($J_{sc}$) is a crucial figure of merit for organic photovoltaic devices, and it is greatly influenced by the intensity and range of the optical absorption spectrum.[31,62,63] The short-circuit current density ($J_{sc}$) is defined as [62,63]

$$J_{sc} = q \int \eta_{EQE}(\lambda) s(\lambda) \, d\lambda \qquad (4)$$

where S($\lambda$) denotes the photon number over all frequencies. The external quantum efficiency ($\eta_{EQE}$) appearing in Eq. (4) can be determined as the product of four physical parameters: the light harvesting efficiency ($\eta_\lambda$), exciton-diffusion efficiency ($\eta_{ED}$), charge transfer efficiency ($\eta_{CT}$) and charge-collection efficiency ($\eta_{CC}$),[62,64] i.e.,

$$\eta_{EQE} = \eta_\lambda \eta_{ED} \eta_{CT} \eta_{CC} \qquad (5)$$

The light harvesting efficiency ($\eta_\lambda$) is associated with the oscillator strength (*f*) of the corresponding excitation with wavelength ($\lambda$) as [31]:

$$\eta_\lambda = 1 - 10^{-f} \qquad (6)$$

The computed optical absorption parameters of the investigated NFAs at the B3LYP-D3BJ/6-31G(d,p) level of theory in the solvent (Chloroform) phase are summarized in Table S6; values are provided for the two highest oscillator strengths values. The simulated optical absorption spectra including the lowest 20 excited states are depicted in Figure 5. About 70% of the solar photon flux is within the wavelength region 380 nm to 900 nm. Therefore, a promising NFA should have strong absorption in this region to harvest more photons, enhancing $J_{sc}$. In photovoltaic cells, both the donor and acceptor contribute to light absorption with photoexcitations generated in their respective domains. The donor PM6 polymer is reported to absorb light in the UV-visible region.[28,65,66] In our study, it is found that all the designed NFAs have the maximum absorption spectra ranging from 597 nm-730 nm, which falls in the visible region to near infra-red (IR) regions of the solar spectrum. The optical absorption spectrum of a molecule strongly depends on the nature of the different substituents in the conjugated backbone. For example, **AQx-2-ct1** has a red-shifted absorption spectrum compared to **AQx-2-c** as the former has more electron withdrawing Cl atoms at its terminal ends. This observation is also valid for **AQx-2-ct2** and **AQx-2-ct3**. In between, the NFAs with thienyl fused IC terminal units, **AQx2-ct4** having γ-thienyl has a smaller optical band gap and hence exhibits a red-shift in its absorption spectrum. Also, in **AQx-2-ct7**, the electron-withdrawing cyano group in the Rhodanine terminal group leads to narrow optical band gap acceptors and displays red-shifted absorption compared to Rhodanine containing **AQx-2-ct6**. Among the designed NFAs, **AQx-2-ct3** shows the largest red-shift having the maximum absorption wavelength of 730 nm as it has an extended π-conjugated terminal unit containing the electron withdrawing -Cl atoms. The NFAs, **AQx-2-ct5**, **AQx-2-ct6**, **AQx-2-ct7**, and **AQx-2-ct8**, have blue-shifted absorption profiles compared to **AQx-2-c** because of their higher band gaps. The short-circuit current density is proportional to the light harvesting efficiency ($\eta_\lambda$). In the present study, the light harvesting efficiency is computed for the brightest excited state transition having the largest oscillator strength. Among all the designed compounds, light harvesting efficiencies for **AQx-2-c**, **AQx-2-ct1**, **AQx-2-ct2**, **AQx-2-ct3**, and **AQx-2-ct4** are slightly higher (~1-2 %) than the other designed NFAs.

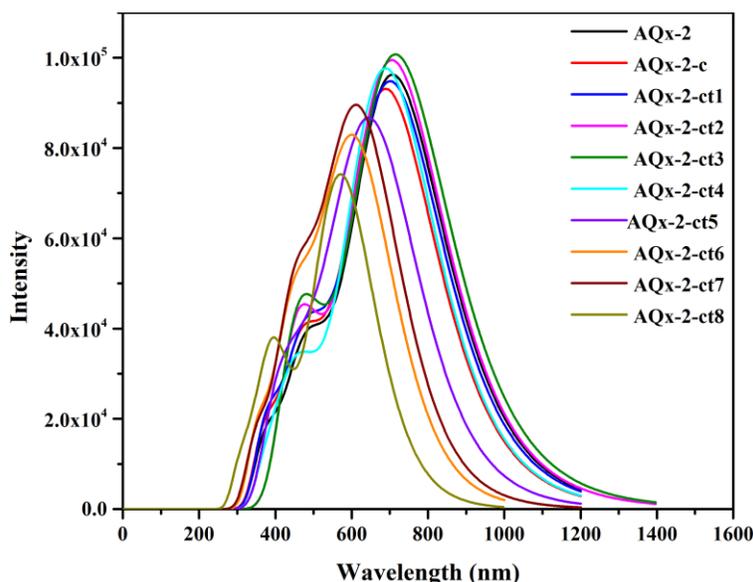

**Figure 5:** Simulated absorption spectra of the designed NFAs in the solution phase (Chloroform) at the B3LYP-D3BJ/6-31G(d,p) level of theory. Spectra include the lowest 20 excitations with Gaussian line shape with peak half-width at half height: 0.333 eV.

**Intramolecular Charge Transport Properties, Electron/Hole Distribution and Transition Density Matrix.**

To investigate the intramolecular charge transport properties of the designed NFAs, the electron/hole distributions of the first bright excited states are computed at the CAM-B3LYP/6-31G(d,p) level of theory level in Chloroform solvent based on gas-phase B3LYP-D3BJ/6-31G(d,p) optimized geometries.[67,68] To understand the electron-hole localization, spatial span, and the primary sites of the singlet excited state electron–hole bound excitons, transition density matrix (TDM) analysis has been performed.[69,70,71] Several different measures of intramolecular charge transport properties are considered: (a) simulated charge density difference ($\triangle\rho$) plots associated with the $S_0 \rightarrow S_1$ transition, (b) the extending zones of the centroids C+ (positive region)/C- (negative region) associated with $S_0 \rightarrow S_1$ transition, and (c) transition density matrix (TDM) analysis. These are illustrated in Figure 6 for the parent **AQx-2-c** compound; results for all the designed NFAs can be found in supplementary information Figures S5-S8. Upon visual inspection, the plots for all the designed NFAs are qualitatively similar to their parent **AQx-2-c**. From the charge density difference (CDD) analysis, the intramolecular charge transfer amounts ($q_{CT}$) of the reference AQx-2 and studied NFAs do not change significantly (0.61-0.67 $|e^-|$) (Table 2), which suggests that the structural modulation of both the core and the terminal units has a small impact on $q_{CT}$. However, the intramolecular charge transition distance ($D_{CT}$) and the variation of dipole moment ($\Delta\mu_{CT}$) between the ground and excited state of the designed NFAs vary significantly. From Table 2, it is clear that the $D_{CT}$ increases slightly after core and terminal unit substitution of the reference NFA AQx-2. The fact is also supported by the computed $\Delta\mu_{CT}$ values. Among the designed NFAs, **AQx-2-ct3** and **AQx-2-ct2** have larger $D_{CT}$ and higher $\Delta\mu_{CT}$ than others due to their longer π-conjugated molecular backbone. This observation also suggests that after local excitation, the intramolecular charge recombination of **AQx-2-ct3** and **AQx-2-ct2** should be more difficult, and charge transfer may be easier than for the other designed NFAs.

The hydrogen atoms have insignificant contribution to the transitions, and, therefore, in our present study, the hydrogen atoms are omitted in Figure 6c and the correlation between the atomic indices of the NFAs and the labels of the TDM plots are shown in Figure S9.[54] The reference NFA and our designed NFAs have the similar $A_2$-$DA_1D$-$A_2$ molecular structures, and, thus, the molecular backbone is subdivided into three parts the terminal electron withdrawing unit ($A_2$), the core unit (D), and the central electron withdrawing part ($A_1$). All the simulated TDM maps show similar behavior where the electron coherence mainly appears at the D parts and the terminal $A_2$ parts. The excitations found in the TDM maps are delocalized over the whole molecular backbone, which is consistent with Figure 6a; compare also Figures S5-S8. The dark sites in these TDM plots are from side alkyl chains, which are not involved in the excitations. From the electron-hole correlation found along the diagonals, during the excitation process, the electrons slightly migrate from the D to the terminal acceptor ($A_2$) part, which is supported by the localization of the LUMO density on terminal units as depicted in Figures S3 and S4.

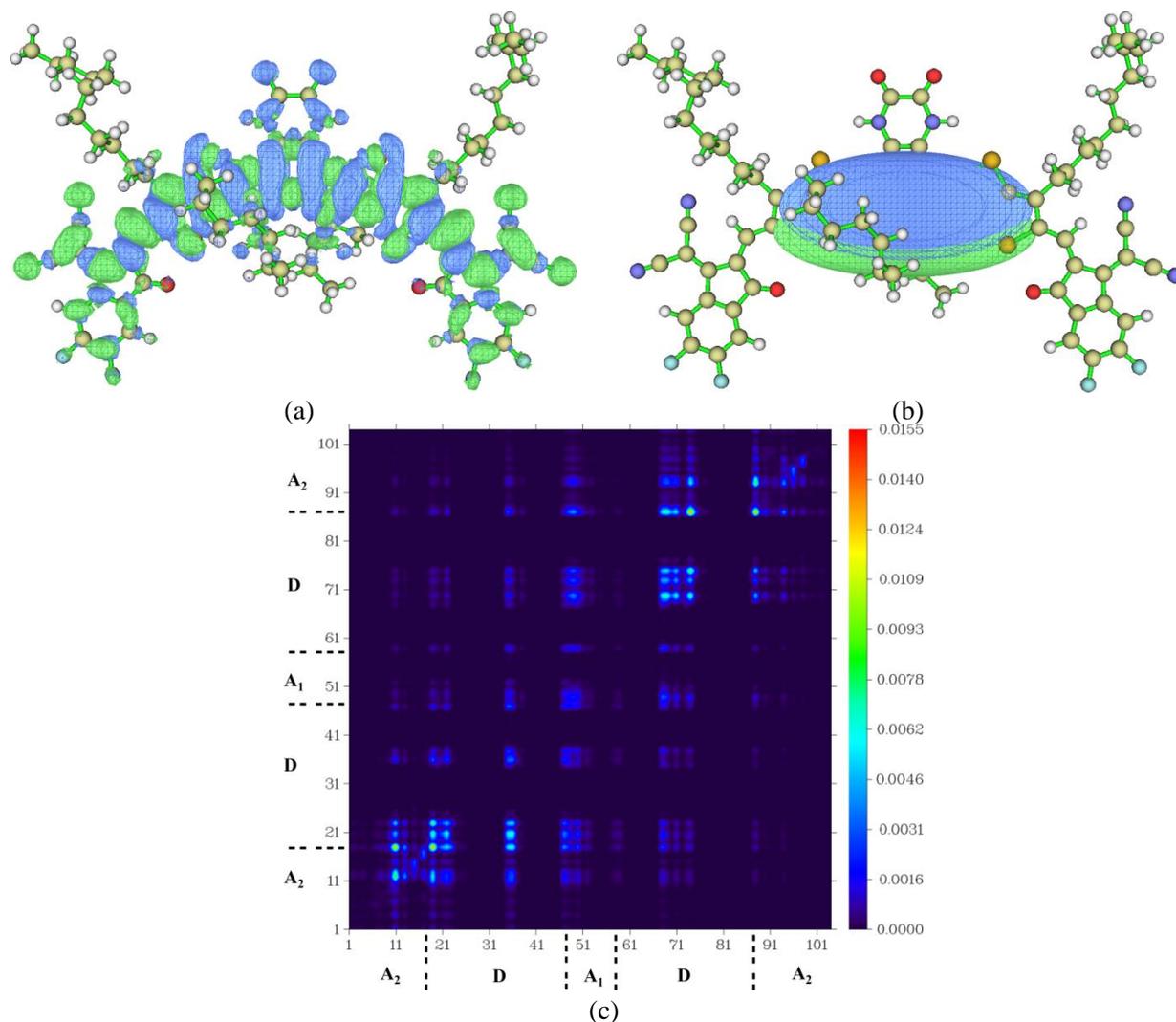

**Figure 6:** Simulated (a) charge density difference (Δρ) plots, (b) extending zones for the centroids C+/C-, and (c) transition density matrix (TDM) associated with the $S_0 \rightarrow S_1$ transition of the studied NFA (**AQx-2-c**) at the CAM-B3LYP/6-31G(d,p) level of theory in Chloroform solvent. The green and blue regions in (a) and (b) correspond to positive and negative regions, respectively and the isosurface value is 0.0002 a.u. In (c), the hydrogen atoms of all molecular systems are omitted and the color bars are given on the right.

**Table 2.** Transferred Charges ($q_{CT}$), Charge Transfer Distance ($D_{CT}$), and Variation of Dipole Moments ($\Delta\mu_{CT}$) for the Transition $S_0 \rightarrow S_1$ at the CAM-B3LYP/6-31G(d,p) Level of Theory in Chloroform Solvent

| Comp. | $q_{CT}$ (\|e⁻\|) | $D_{CT}$ (Å) | $\Delta\mu_{CT}$ (Debye) |
|---|---|---|---|
| AQx-2 (ref) | 0.65 | 1.137 | 3.564 |
| **AQx-2-c** | 0.66 | 1.249 | 3.993 |
| **AQx-2-ct1** | 0.67 | 1.317 | 4.226 |
| **AQx-2-ct2** | 0.67 | 1.410 | 4.523 |
| **AQx-2-ct3** | 0.67 | 1.484 | 4.777 |
| **AQx-2-ct4** | 0.66 | 1.337 | 4.215 |
| **AQx-2-ct5** | 0.67 | 1.193 | 3.859 |
| **AQx-2-ct6** | 0.63 | 1.380 | 4.189 |
| **AQx-2-ct7** | 0.64 | 1.362 | 4.233 |
| **AQx-2-ct8** | 0.61 | 1.159 | 3.410 |

## Electron Affinity (EA), Ionization Potential (IP), Mulliken electronegativity (ME) and Reorganization energy.

The ionization potentials (IPs), electron affinities (EAs), and reorganization energies play important roles in the charge transport mechanism in organic molecules. The adiabatic and vertical ionization potentials, as well as electron affinities, are listed in Table 3. The computed values of adiabatic EA for all the designed NFAs except **AQx-2-ct6**, are found to be higher than 2.4 eV, suggesting their air-stability and n-type characteristics.[72,73] The fact is also supported by the shallow LUMO energy of **AQx-2-ct6**, as discussed in electronic properties and air-stability Section. The computed values of the Mulliken Electronegativity (ME) are also listed in Table 3. The Mulliken electronegativity is calculated as half of the sum of the vertical ionization potential (IP(v)) and the vertical electron affinity (EA(v)) energy. A molecule having a larger Mulliken electronegativity index possesses a higher electron withdrawing ability[74], and, thus, all the designed NFAs, except **AQx-2-ct5, AQx-2-ct6**, and **AQx-2-ct7**, have larger electron withdrawing ability than the reference AQx-2. In the present study, the contribution of the external reorganization energy ($\lambda_{ext}$) is neglected as the contribution of $\lambda_{ext}$ is expected to be less sensitive to the chemical structure.[75,76,77] The internal reorganization ($\lambda_i$) energies of all the investigated NFAs during hole and electron transfer are evaluated using the adiabatic potential energy surface (APS) method as described in Section S.2.3 in the supporting information.[78] From Table 3, the electron reorganization energy ($\lambda_e$) is significantly lower than the hole reorganization energy ($\lambda_h$) for all the designed NFAs except for **AQx-2-ct6** and **AQx-2-ct7**. The NFAs having lower $\lambda_e$ than $\lambda_h$ should be better for electron transport. For **AQx-2-ct8**, both values of $\lambda_e$ and $\lambda_h$ are almost the same, and, hence, it may have balanced hole and electron transport rates. Moreover, the charge transport efficiency also depends on the charge transfer integral, which is influenced by molecular packing.

**Table 3.** Hole/electron Reorganization energies ($\lambda_h/\lambda_e$), Adiabatic/Vertical Ionization Potential (IP(a)/IP(v)), Adiabatic/Vertical Electron Affinity (EA(a)/ EA(v)) and Mulliken Electronegativity (ME) in eV of the studied NFAs

| Comp. | $\lambda_h$ | $\lambda_e$ | IP(a) | IP(v) | EA(a) | EA(v) | ME |
|---|---|---|---|---|---|---|---|
| AQx-2 (Ref.) | 0.176 | 0.132 | 6.22 | 6.31 | 2.77 | 2.70 | 4.50 |
| **AQx-2-c** | 0.177 | 0.132 | 6.42 | 6.51 | 2.92 | 2.85 | 4.68 |
| **AQx-2-ct1** | 0.175 | 0.122 | 6.46 | 6.54 | 3.01 | 2.95 | 4.74 |
| **AQx-2-ct2** | 0.170 | 0.115 | 6.33 | 6.41 | 2.94 | 2.88 | 4.65 |
| **AQx-2-ct3** | 0.168 | 0.106 | 6.37 | 6.46 | 3.02 | 2.97 | 4.72 |
| **AQx-2-ct4** | 0.168 | 0.127 | 6.34 | 6.43 | 2.84 | 2.78 | 4.60 |
| **AQx-2-ct5** | 0.183 | 0.137 | 6.27 | 6.36 | 2.64 | 2.58 | 4.47 |
| **AQx-2-ct6** | 0.182 | 0.197 | 6.02 | 6.11 | 2.24 | 2.14 | 4.13 |
| **AQx-2-ct7** | 0.172 | 0.204 | 6.26 | 6.34 | 2.56 | 2.46 | 4.40 |
| **AQx-2-ct8** | 0.166 | 0.170 | 6.60 | 6.68 | 2.49 | 2.41 | 4.54 |

## Crystal Structure Prediction, Charge Transfer Integral, and Anisotropic Mobility.

For every NFA, including the reference compound AQx-2, the structure with the lowest lattice energy from the considered space groups has been selected for further calculations. The simulated unit cell parameters of the lowest energy space group for all the NFAs are given in Table S7. The simulation details are mentioned in Section S1 in the Supporting Information. The packing motifs of all the investigated NFAs are found to be π-stacked parallel, where example are shown for AQx-2, **AQx-2-c** and **AQx-2-ct3** in Figure 7; all other packing structures can be found in SI (Figure S10-S12). Based on the predicted crystal structures, we consider one molecule in a crystal as the center, and its adjacent parallel dimers are represented as D1-D8. The rate of charge transfer (K) are calculated using Marcus theory (as described in Section S.2.1 in the Supporting Information) along different pathways, as depicted in Figures 7 and S10-S12.[78,79,80] The charge transfer integral or intermolecular electron coupling (V) is one of the key parameters influencing the charge carrier mobility, and it is determined by the orbital coupling of the adjacent molecules as described in Section 2.2.2 in the Supporting Information.[46,47,81] The degree of orbital coupling is affected by the different packing motifs and the relative orientation of the nearest neighbor molecules.

The computed charge transport parameters along different hopping dimers and charge carrier mobilities of the studied NFAs are listed in Table 4 and Table 5. In organic crystals, different molecular packing motifs along dissimilar hopping pathways and the variation in the charge transfer integrals of the central molecule with the adjacent molecules produces anisotropic mobility.[82,83] The anisotropic mobility of the studied NFAs are evaluated using the formulation as described in Section S.2.4 in the Supporting information. The anisotropic electron/hole mobilities ($\mu_e/\mu_h$) of a subset of the studied NFAs are shown in Figure 7, while the plots for all other NFAs are available in Figure S10-S12 in the Supporting Information. The calculated mobilities are listed in Table 4 and Table 5. The maximum values of $\mu_e$ and $\mu_h$ in **AQx-2-c** are observed to be $5.47 \times 10^{-1}$ cm$^2$V$^{-1}$s$^{-1}$ and $4.15 \times 10^{-4}$ cm$^2$V$^{-1}$s$^{-1}$, respectively, corresponding to $\phi$ = 1.2°/0.0°. Therefore, in **AQx-2-c**, the electron mobility is found to be much higher than the hole mobility suggesting its electron-transporting nature, and this observation is also supported by the higher electron transfer integrals (10.66 meV) and higher electron transfer rate ($1.4616 \times 10^{12}$ s$^{-1}$) along the same hopping channel. Among the designed NFAs, **AQx-2-ct3** exhibits the highest electron mobility of 6.03 cm$^2$V$^{-1}$s$^{-1}$, obtained along $\phi$ = 38.3°/218.3°. All the designed NFAs have better charge carrier mobilities than the reference AQx-2. Hence, the designed NFAs may act as better electron transporting acceptor materials in organic photovoltaic applications. It is also observed that all the designed NFAs except **AQx-2-ct7** have better electron mobility than hole mobility, and hence, it suggests that our designed NFAs are promising electron transporting acceptor materials.

**Table 4.** The Intermolecular Distances (Measured Distance Between Their Mass-Centers) Along Different Pathways, Angle of Orientation ($\theta$), Hole/Electron Couplings ($V_h$/ $V_e$), Transfer Rates of Hole/Electron ($K_h$/$K_e$), and Angle-Resolved Hole/Electron Mobility Ranges ($\mu_h$/$\mu_e$) of the Designed NFAs

| Comp. | Pathways | r (Å) | $\theta$ (deg) | $V_h$ (meV) | $V_e$ (meV) | $K_h$ (s$^{-1}$) | $K_e$ (s$^{-1}$) | $\mu_h$(cm$^{-1}$V$^{-1}$s$^{-1}$) | $\mu_e$(cm$^{-1}$V$^{-1}$s$^{-1}$) | $\mu_e/\mu_h$ |
|---|---|---|---|---|---|---|---|---|---|---|
| AQx-2 (Ref.) | D1, D5 | 18.682 | 0 | 0.02 | 0.12 | 2.9116×10$^6$ | 1.8522×10$^8$ | 0-1.97×10$^{-6}$ | 0-1.25×10$^{-4}$ | 6.34×10$^1$ |
|  | D2, D6 | 22.628 | 34.35 | 0.00 | 0.00 | 0 | 0 |  |  |  |
|  | D3, D7 | 12.767 | 90.00 | 0.00 | 0.00 | 0 | 0 |  |  |  |
|  | D4, D8 | 22.628 | 145.65 | 0.00 | 0.00 | 0 | 0 |  |  |  |
| **AQx-2-c** | D1, D5 | 14.770 | 0 | 0.37 | 10.66 | 9.8410×10$^8$ | 1.4616×10$^{12}$ | 1.79×10$^{-10}$-4.15×10$^{-4}$ | 9.83×10$^{-3}$-5.47×10$^{-1}$ | 1.32×10$^3$ |
|  | D2, D6 | 19.795 | 42.20 | 0.01 | 4.16 | 7.1885×10$^5$ | 2.2259×10$^{11}$ |  |  |  |
|  | D3, D7 | 13.296 | 90.45 | 0.00 | 0.01 | 0 | 1.2863×10$^6$ |  |  |  |
|  | D4, D8 | 19.951 | 138.21 | 0.00 | 0.00 | 0 | 0 |  |  |  |
| **AQx-2-ct1** | D1, D5 | 15.033 | 0 | 0.69 | 4.23 | 3.5091×10$^9$ | 2.6370×10$^{11}$ | 1.34×10$^{-7}$-1.52×10$^{-3}$ | 1.09×10$^{-2}$-6.27×10$^{-1}$ | 4.12×10$^2$ |
|  | D2, D6 | 20.326 | 42.54 | 0.00 | 0.00 | 0 | 0 |  |  |  |
|  | D3, D7 | 13.743 | 90.24 | 0.07 | 0.04 | 3.6116×10$^7$ | 2.3580×10$^7$ |  |  |  |
|  | D4, D8 | 20.410 | 137.67 | 0.01 | 8.11 | 7.371×10$^5$ | 9.6934×10$^{11}$ |  |  |  |
| **AQx-2-ct2** | D1, D5 | 15.323 | 0 | 0.29 | 0.10 | 6.6007×10$^8$ | 1.6243×10$^8$ | 0-3.00×10$^{-4}$ | 1.71×10$^{-7}$-1.23×10$^{-1}$ | 4.10 × 10$^2$ |
|  | D2, D6 | 26.438 | 66.96 | 0.00 | 0.00 | 0 | 0 |  |  |  |
|  | D3, D7 | 24.834 | 101.56 | 0.00 | 2.52 | 0 | 1.0315×10$^{11}$ |  |  |  |
|  | D4, D8 | 31.687 | 129.84 | 0.00 | 0.01 | 0 | 1.6243×10$^6$ |  |  |  |
| **AQx-2-ct3** | D1, D5 | 16.207 | 0 | 0.07 | 0.14 | 3.944×10$^7$ | 3.6175×10$^8$ | 1.11×10$^{-8}$ -1.93×10$^{-5}$ | 2.94×10$^{-6}$-6.0262 | 3.12 × 10$^5$ |
|  | D2, D6 | 21.809 | 38.26 | 0.01 | 18.84 | 8.049 × 10$^5$ | 6.5512 × 10$^{12}$ |  |  |  |
|  | D3, D7 | 13.537 | 86.12 | 0.01 | 0.01 | 8.049 × 10$^5$ | 1.8457 × 10$^6$ |  |  |  |
|  | D4, D8 | 20.401 | 138.55 | 0.00 | 0.00 | 0 | 0 |  |  |  |

**Table 5.** The Intermolecular Distances (Measured Distance Between Their Mass-Centers) Along Different Pathways, Angle of Orientation (θ), Hole/Electron Couplings ($V_h$/ $V_e$), Transfer Rates of Hole/Electron ($K_h$/$K_e$), and Angle-Resolved Hole/Electron Mobility Ranges ($\mu_h$/$\mu_e$) of the

| Comp. | Pathways | r (Å) | θ (deg) | $V_h$ (meV) | $V_e$ (meV) | $K_h$ ($s^{-1}$) | $K_e$ ($s^{-1}$) | $\mu_h$ ($cm^{-1}V^{-1}s^{-1}$) | $\mu_e$ ($cm^{-1}V^{-1}s^{-1}$) | $\mu_e/\mu_h$ |
|---|---|---|---|---|---|---|---|---|---|---|
| **AQx-2-ct4** | D1, D5 | 13.208 | 0 | 0.79 | 2.33 | 5.0237×10$^9$ | 7.4718×10$^{10}$ | 6.17×10$^{-5}$-3.04×10$^{-2}$ | 6.42×10$^{-4}$-6.72×10$^{-1}$ | 2.20×10$^1$ |
| | D2, D6 | 14.964 | 49.56 | 3.06 | 10.86 | 7.5372×10$^{10}$ | 1.6232×10$^{12}$ | | | |
| | D3, D7 | 11.915 | 107.09 | 0.27 | 0.06 | 5.8681×10$^8$ | 4.9547×10$^7$ | | | |
| | D4, D8 | 20.222 | 145.72 | 0.00 | 0.00 | 0 | 0 | | | |
| **AQx-2-ct5** | D1, D5 | 19.211 | 0 | 0.03 | 3.28 | 6.0040×10$^6$ | 1.2942×10$^{11}$ | 3.54×10$^{-7}$-3.87×10$^{-5}$ | 1.14×10$^{-2}$-7.84×10$^{-2}$ | 2.03×10$^3$ |
| | D2, D6 | 26.378 | 43.26 | 0.00 | 2.44 | 0 | 7.1620×10$^{10}$ | | | |
| | D3, D7 | 18.076 | 90.00 | 0.10 | 0.89 | 6.6711×10$^7$ | 9.5287×10$^9$ | | | |
| | D4, D8 | 26.378 | 136.74 | 0.00 | 0.00 | 0 | 0 | | | |
| **AQx-2-ct6** | D1, D5 | 18.287 | 0 | 0.53 | 10.36 | 1.8973×10$^9$ | 6.0272×10$^{11}$ | 7.85×10$^{-11}$-1.23×10$^{-3}$ | 0-3.89×10$^{-1}$ | 3.16×10$^2$ |
| | D2, D6 | 18.490 | 44.64 | 0.01 | 0.00 | 6.7544×10$^5$ | 0 | | | |
| | D3, D7 | 13.967 | 111.55 | 0.00 | 0.00 | 0 | 0 | | | |
| | D4, D8 | 26.779 | 150.98 | 0.00 | 0.00 | 0 | 0 | | | |
| **AQx-2-ct7** | D1, D5 | 20.071 | 0 | 3.47 | 2.24 | 9.2155×10$^{10}$ | 2.5877×10$^{10}$ | 3.64×10$^{-3}$-8.71×10$^{-1}$ | 1.732×10$^{-3}$-1.212×10$^{-1}$ | 1.39×10$^{-1}$ |
| | D2, D6 | 25.108 | 44.44 | 10.19 | 4.81 | 7.9470×10$^{11}$ | 1.1932×10$^{11}$ | | | |
| | D3, D7 | 17.710 | 96.95 | 0.45 | 0.19 | 1.5498×10$^9$ | 1.8617×10$^8$ | | | |
| | D4, D8 | 28.329 | 141.64 | 0.00 | 0 | 0 | 0 | | | |
| **AQx-2-ct8** | D1, D5 | 13.453 | 0 | 0.09 | 0.02 | 6.6873×10$^7$ | 3.1395×10$^6$ | 8.85×10$^{-6}$-1.27×10$^{-5}$ | 1.52×10$^{-9}$-5.05×10$^{-4}$ | 3.97×10$^1$ |
| | D2, D6 | 17.917 | 34.84 | 0.00 | 0.00 | 0 | 0 | | | |
| | D3, D7 | 10.313 | 83.03 | 0.11 | 0.56 | 9.9897×10$^7$ | 2.4613×10$^9$ | | | |
| | D4, D8 | 15.927 | 140.00 | 0.01 | 0.00 | 8.2559×10$^5$ | 0 | | | |

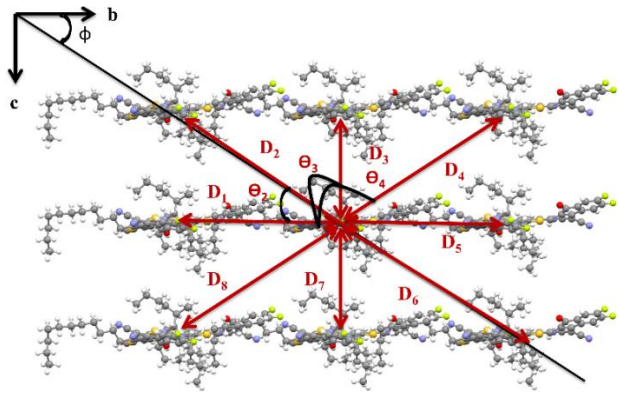
(a) AQx-2

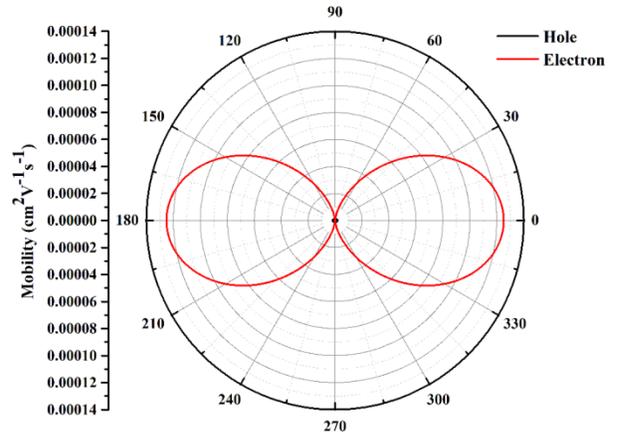
(b) AQx-2

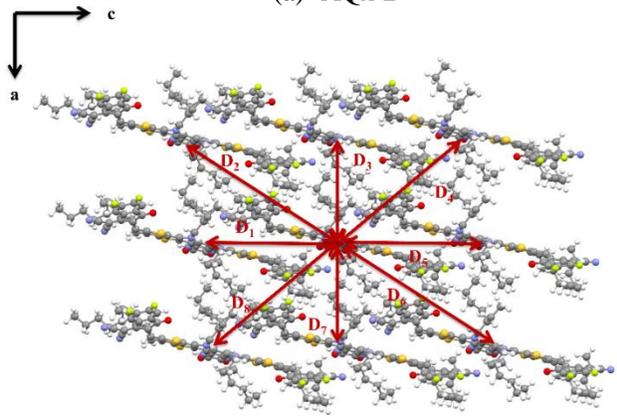
(c) **AQx-2-c**

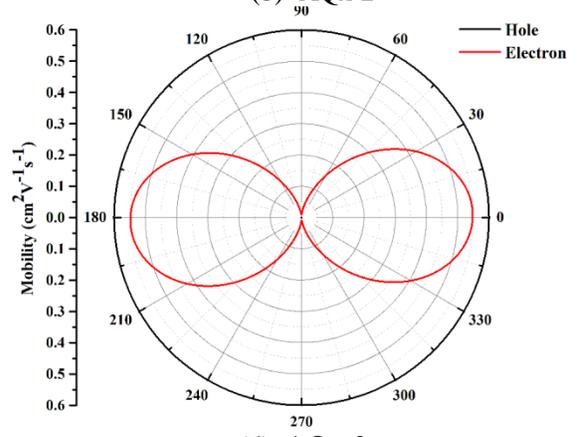
(d) **AQx-2-c**

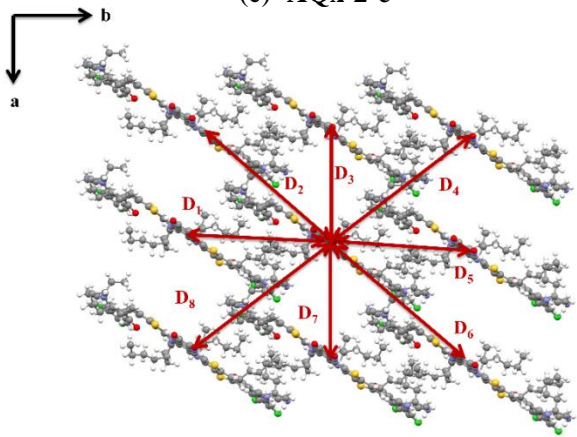
(e) **AQx-2-ct3**

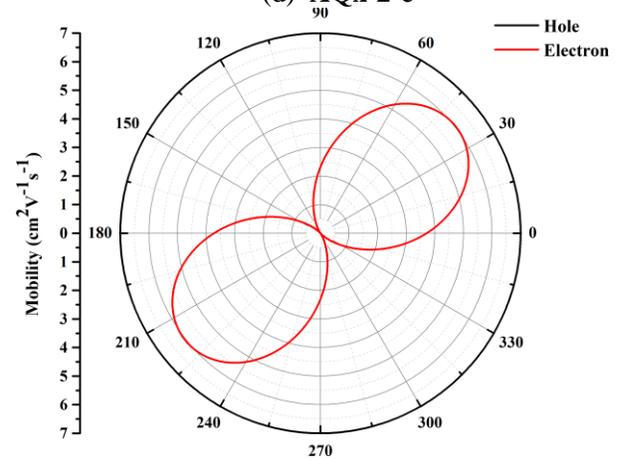
(f) **AQx-2-ct3**

**Figure 7:** Simulated crystal structures showing different hopping channels and angle resolved mobility of the studied NFAs (AQx-2, **AQx-2-c**, **AQx-2-ct3**).

## Absorption Properties and Charge Transfer Analysis of the D/A Blends.

The optical absorption of D/A blends and the charge transfer mechanism from donor to acceptor play important roles in the organic photovoltaic performance. The blends of the PM6 donor and designed NFAs are built by A-A stacking. The A-A stacking of such D/A blends was widely accepted because such a stacking pattern seems to be dominant and favorable for exciton dissociation and charge transfer.[2,84] In the complex systems of the designed NFAs and the PM6 donor, A-A stacking involves the terminal electron-withdrawing end groups of the NFAs and the electron-withdrawing unit of the polymer donor. To simplify the calculations, in our study, only one repeat unit of the polymer donor PM6 was taken into account, and unsaturated atoms were hydrogenated. Initially, the distance between the terminal electron-withdrawing end groups of the NFAs and the electron-withdrawing unit of the PM6 has been set as 3.5 Å (Figure 8). Further, we optimized the D/A (PM6:NFA)) blend structure using the cost-effective extended tight-binding (GFN2-xTB) method.[51] To analyze the optical properties and charge transfer behaviour of the PM6:NFA blends, TD-DFT calculations are carried out at the CAM-B3LYP/6-31G(d,p) level of theory in the solvent (Chloroform) phase based on GFN2-xTB optimized geometries. It is observed from Table 6 and Figure 9 that the maximum absorption wavelengths (503-610 nm) of the PM6:NFA blends lie in the visible region of the solar spectrum. The light harvesting efficiencies ($\eta_\lambda$s) of the blends are computed corresponding to the brightest excited state transition. The computed light harvesting efficiencies lie in the range of 0.9589-0.9957, which indicates efficient light harvesting by the PM6:NFA blends during solar cell device operation. Among the designed blends, PM6:**AQx-2-ct6** and PM6:**AQx-2-ct8** have slightly smaller light harvesting efficiencies compared to the other blends and, therefore, the other PM6:NFA blends may have better short-circuit current density values. The intermolecular charge transfer (CT) states of the PM6:NFA blends are identified by natural transition orbital (NTO) analysis considering the first 40 excited states, as shown for AQx-2, **AQx-2-c** and **AQx-2-ct3** in Figure 10 and all the other NFAs are given in Figures S13-S14 in the Supporting Information. In the CT state, the HOMO predominantly localizes on the donor PM6 and the LUMO localizes on the acceptor NFA. The excited state properties of the corresponding CT states, along with NTO contributions, are listed in Table S8 in the Supporting Information. In addition, the $\Delta r$ parameter is also evaluated for the listed excited states to verify their CT nature (Table S8). The computed values of the $\Delta r$ parameter are found to be larger than 2.0 Å (6.2-15.4 Å), which indicates that the identified excited states are mainly charge transfer (CT) states.[54,85] We also carried out interfragment charge transfer (IFCT) analysis to validate the nature of the excitation and to quantify the net electron transfer from the donor PM6 fragment to acceptor fragments. The contribution of each fragment to the hole and electron as well as the percentage of transferred electrons between the donor and acceptor fragments are listed in Table 7. The percentage of intrafragment charge transfer is improved for the designed PM6:NFA blends (except PM6:**AQx-2-ct5**) compared to the PM6:AQx-2 (reference NFA) blend. Thus, for those designed PM6:NFA blends, the interfacial exciton dissociation process may be easier than the previously reported

PM6 and AQx-2 blend and the newly designed PM6:NFA blends may promote better charge carrier separation at the D/A interface.

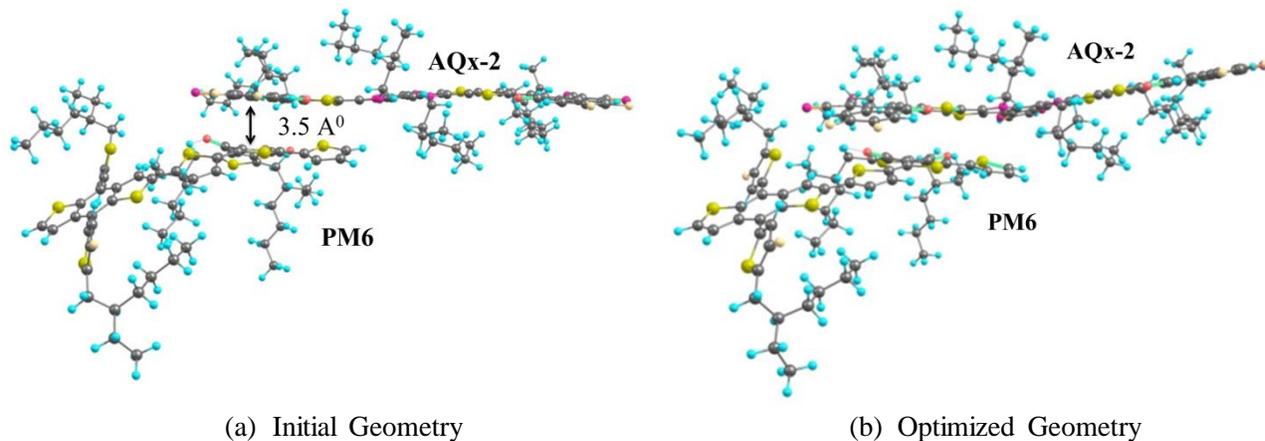

(a) Initial Geometry  (b) Optimized Geometry

**Figure 8.** The initial and optimized geometries of the investigated PM6/NFAs blends (Representative blend: PM6:AQx-2).

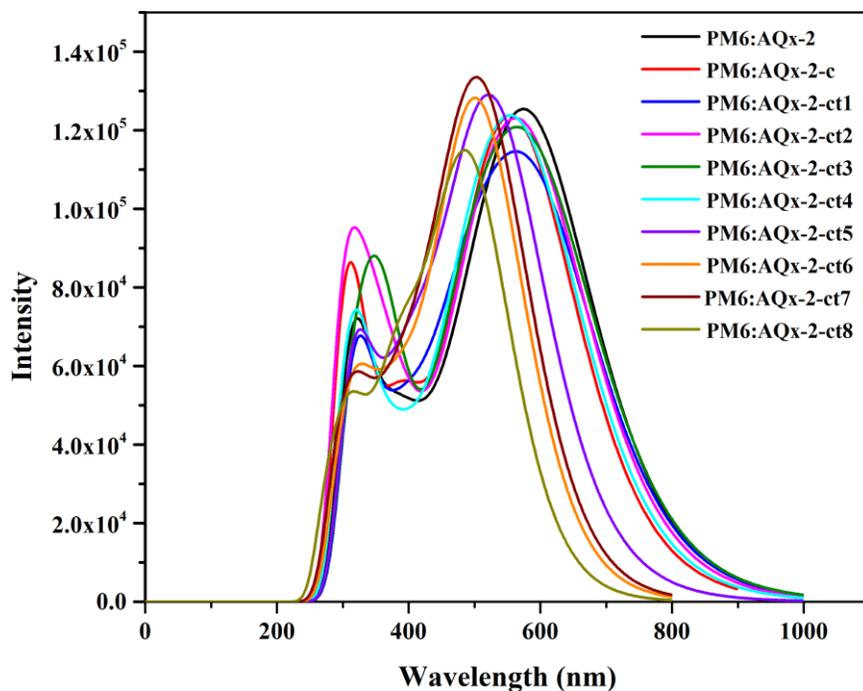

**Figure 9:** The optical absorption spectra of the investigated PM6/NFAs complexes at the CAM-B3LYP/6-31G(d,p) level of theory in the solvent (Chloroform) phase. Spectra include the lowest 40 excitations with Gaussian lineshape with peak half-width at half height: 0.333 eV.

**Table 6.** Computed Electronic Transition Energy E (eV), Absorption Wavelength $\lambda_{abs}$ (nm), Major Configurations, Oscillator Strength (*f*), and Light Harvesting Efficiency ($\eta_\lambda$) of the PM6/NFA Blends as Determined at the CAM-B3LYP/6-31G(d,p) Level of Theory in the Solvent (Chloroform) Phase

| Complex | State | E (eV) | $\lambda_{abs}$ (nm) | major config. | *f* | $\eta_\lambda$ |
|---|---|---|---|---|---|---|
| PM6:AQx-2 | $S_0 \to S_1$ | 2.05 | 604 | H-1 → L (56%) | 2.3215 | 0.9952 |
| | $S_0 \to S_3$ | 2.51 | 495 | H-1 → L+1 (51%) | 0.6992 | |
| **PM6:AQx-2-c** | $S_0 \to S_1$ | 2.11 | 588 | H-1 → L (75%) | 2.2281 | 0.9941 |
| | $S_0 \to S_3$ | 2.46 | 503 | H → L+2 (37%) | 0.7841 | |
| **PM6:AQx-2-ct1** | $S_0 \to S_1$ | 2.03 | 609 | H-1 → L (49%) | 2.0369 | 0.9908 |
| | $S_0 \to S_4$ | 2.55 | 486 | H → L+3 (42%) | 0.9240 | |
| **PM6:AQx-2-ct2** | $S_0 \to S_1$ | 2.06 | 601 | H-1 → L (75%) | 2.2259 | 0.9941 |
| | $S_0 \to S_3$ | 2.45 | 507 | H-1 → L+1 (44%) | 0.7357 | |
| **PM6:AQx-2-ct3** | $S_0 \to S_1$ | 2.03 | 610 | H-1 → L (73%) | 2.1759 | 0.9933 |
| | $S_0 \to S_4$ | 2.54 | 488 | H → L+2 (54%) | 0.9148 | |
| **PM6:AQx-2-ct4** | $S_0 \to S_1$ | 2.09 | 593 | H-1 → L (50%) | 2.2159 | 0.9939 |
| | $S_0 \to S_4$ | 2.55 | 485 | H → L+2 (36%) | 0.9035 | |
| **PM6:AQx-2-ct5** | $S_0 \to S_1$ | 2.06 | 602 | H → L (61%) | 0.2734 | |
| | $S_0 \to S_2$ | 2.30 | 538 | H-1 → L (42%) | 2.3658 | 0.9957 |
| | $S_0 \to S_4$ | 2.60 | 478 | H → L+3 (38%) | 0.6087 | |
| **PM6:AQx-2-ct6** | $S_0 \to S_1$ | 2.37 | 523 | H → L (75%) | 1.6703 | 0.9786 |
| | $S_0 \to S_2$ | 2.47 | 503 | H-1 → L+1 (66%) | 1.0402 | |
| **PM6:AQx-2-ct7** | $S_0 \to S_1$ | 2.35 | 528 | H-1 → L (61%) | 2.1199 | 0.9924 |
| | $S_0 \to S_3$ | 2.71 | 457 | H-1 → L+1 (34%) | 1.0586 | |
| **PM6:AQx-2-ct8** | $S_0 \to S_1$ | 2.46 | 503 | H → L+2 (34%) | 1.3863 | 0.9589 |
| | $S_0 \to S_2$ | 2.54 | 488 | H-1 → L (80%) | 1.1781 | |

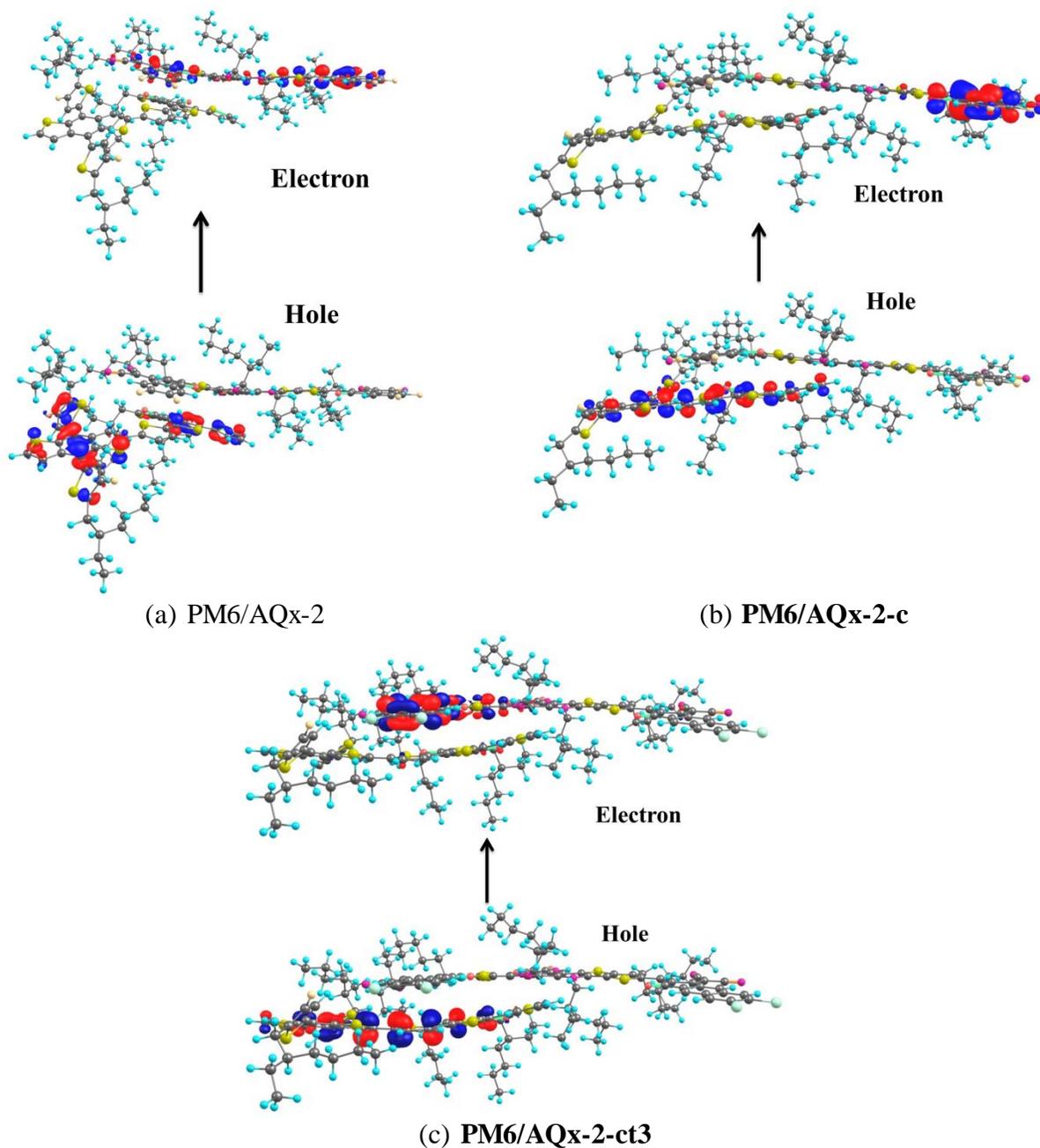

(a) PM6/AQx-2  (b) **PM6/AQx-2-c**

(c) **PM6/AQx-2-ct3**

**Figure 10:** The NTO analysis of the CT excited states of the investigated PM6/NFA (AQx-2, **AQx-2-c**, **AQx-2-ct3**) blends. The blue and red colors represent the positive and negative isosurfaces, respectively.

**Table 7.** Contribution of Donor, Acceptor Fragments to Hole and Electron and Percentage of Interfragment Charge Transfer (IFCT) Between Donor, Acceptor Fragments in CT States

| Complex | CT state | Hole (Electron) contribution of PM6 | Hole (Electron) contribution of NFAs | IFCT (PM6 → NFA) | IFCT (NFA → PM6) |
|---|---|---|---|---|---|
| PM6:AQx-2 | S28 | 71.54% (5.27%) | 28.46% (94.73%) | 67.76% | 1.50% |
| **PM6:AQx-2-c** | S38 | 96.64% (0.97%) | 3.36% (99.03%) | 95.70% | 0.03% |
| **PM6:AQx-2-ct1** | S30 | 92.61% (22.76%) | 7.39% (77.24%) | 71.53% | 1.68% |
| **PM6:AQx-2-ct2** | S29 | 90.08% (8.30%) | 9.92% (91.70%) | 82.60% | 0.82% |
| **PM6:AQx-2-ct3** | S33 | 91.84% (24.88%) | 8.16% (75.12%) | 68.99% | 2.03% |
| **PM6:AQx-2-ct4** | S30 | 86.91% (9.90%) | 13.09% (90.10%) | 78.30% | 1.29% |
| **PM6:AQx-2-ct5** | S38 | 69.53% (12.33%) | 30.47% (87.67%) | 60.95% | 3.76% |
| **PM6:AQx-2-ct6** | S23 | 77.70% (2.98%) | 22.30% (97.02%) | 75.39% | 0.66% |
| **PM6:AQx-2-ct7** | S19 | 91.70% (4.85%) | 8.30% (95.15%) | 87.25% | 0.40% |
| **PM6:AQx-2-ct8** | S30 | 99.62 % (15.10%) | 0.38 % (84.90%) | 84.58% | 0.06% |

# CONCLUSION

In the present study, we theoretically designed a series of non-fullerene acceptors (NFAs) based on the previously reported crescent shaped AQx-2. At first, we performed core modulation of the reported AQx-2 and designed a new NFA, namely **AQx-2-c**, in which the quinoxaline core of AQx-2 is replaced by 2,3-quinoxalinedione,1,4-dihydro having a weaker electron donating ability than quinoxaline. Further, we designed a series of eight more NFAs by performing terminal-unit engineering of **AQx-2-c** (terminal unit: Fluorinated IC) with various halogen substituted, π-extended, thienyl fused IC groups and different end-capped groups other than IC also. We explored the geometrical, optoelectronic, intra and intermolecular charge transport properties, and overall photovoltaic performance of the designed crescent shaped NFAs with respect to the PM6 donor. From the geometrical analysis, we observed that all the designed NFAs except **AQx-2-ct8** have near planar conjugated backbones as their dihedral angles do not vary significantly from zero. All the computed C-C bond lengths exhibited significant double bond character, which facilitates the delocalization of π-electrons over the whole molecular backbone and improves the charge transport properties. From frontier molecular orbital analysis, the computed HOMO levels of the investigated designed NFAs were below the air oxidation threshold (ca. -5.27 eV or 0.57 V vs saturated calomel electrode (SCE)) and found to be air-stable. Except for **AQx-2-ct6**, all the designed NFAs could have more n-type character as their computed LUMO levels were found to be close to -4.0 eV and their adiabatic electron affinity values were observed to be higher than 2.4 eV. This observation was also supported by the significantly lower electron injection barriers of the studied NFAs other than for **AQx-2-ct6**. The designed NFAs except for **AQx-2-ct5**, **AQx-2-ct6**, and **AQx-2-ct7** may favor better electron injection than the reported AQx-2 as they had smaller electron injection barriers due to their lower-lying LUMO levels. The studied NFAs could act as potential non-fullerene acceptor candidates as they were found to have sufficient open-circuit voltages and fill factors ranging from 0.62-1.41 eV and 83%-91%, respectively. All the newly designed NFAs except **AQx-2-ct5**, **AQx-2-ct6**, and **AQx-2-ct7**, were expected to have larger electron withdrawing abilities than the reported AQx-2 as they have higher Mulliken electronegativity indices than AQx-2. From the anisotropic mobility analysis, it was noted that all the designed NFAs have better charge carrier mobilities than the reference AQx-2. It was also found that all the designed NFAs except **AQx-2-ct7** had better electron mobilities than hole mobilities, and hence, it suggests that our designed NFAs could perform as good electron transport acceptor materials in organic solar cells. The charge density difference (CDD) analysis suggested that the structural modulation of the core and terminal units had only a small impact on intramolecular charge transfer amounts ($q_{CT}$) as the computed values of $q_{CT}$ for the reference AQx-2 and the studied NFAs did not change significantly (0.63-0.67 |e⁻|). However, **AQx-2-ct2** and **AQx-2-ct3**, having longer π-conjugated terminal units, showed relatively longer $D_{CT}$ than the other NFAs, suggesting more difficult intramolecular charge recombination and easier charge transfer for these two compounds. The excitations found in the TDM map were delocalized over the whole molecular backbone, and during the excitation process, the electrons were found to be slightly migrated from the D to the terminal acceptor ($A_2$) part. From the optical absorption analysis, all the designed NFAs had the maximum absorption spectra ranging from 597 nm-730 nm, which lay in the visible and near infra-red (IR) regions of the solar spectrum. Among the studied NFAs, the absorption profile of **AQx-2-ct3** showed the maximum red-shift (730 nm) due to its extended π-conjugated terminal unit containing the electron withdrawing -Cl atoms. The computed light harvesting efficiencies for **AQx-2-c**, **AQx-2-ct1**, **AQx-2-ct2**, **AQx-2-ct3**, and **AQx-2-ct4** were found to be slightly higher than the other designed NFAs and they may exhibit better short-circuit current density. The simulated maximum absorption wavelengths (503-610 nm) of the PM6:NFA blends were observed to lie in the visible region of the solar spectrum. The computed light harvesting efficiencies for the D/A blends were found to be in the range of 0.9589-0.9957, which indicates efficient light harvesting by the PM6:NFA blends during solar cell device operation. Among the designed blends, PM6:**AQx-2-ct6** and PM6:**AQx-2-ct8** have slightly smaller light harvesting efficiencies compared to the other blends, and therefore, the other PM6:NFA blends may have better short-circuit current density values. From the intrafragment charge transfer (IFCT) analysis, it was observed that the amount of intrafragment charge transfer is improved for the designed PM6:NFA blends (except PM6:**AQx-2-ct5**) over the PM6:AQx-2 (reference NFA) blend. The fact indicates that, for our designed PM6:NFA blends, the interfacial exciton dissociation process may be easier than the previously reported PM6:AQx-2 blend, and the newly designed PM6:NFA blends may promote better charge carrier separation at the D/A interface. It is also noted that among all the newly designed NFAs, **AQx-2-ct3** may be the best choice as an acceptor material. **AQx-2-ct3** had the lowest optical band gap (1.70 eV) along with the maximum red-shifted optical absorption (730 nm) as it had an extended π-conjugated terminal unit containing the electron withdrawing -Cl atoms. The lowest electron injection barrier and lowest electron reorganization energy led to the highest electron mobility 6.03 $cm^2V^{-1}s^{-1}$ of **AQx-2-ct3**. In addition, **AQx-2-ct3** had better ICT properties than other designed NFAs as it had the highest $D_{CT}$ and $\Delta\mu_{CT}$.

In summary, our theoretical investigation showed a series of promising non-fullerene acceptors, and we expect that our study will provide further insight into the theoretical basis for the synthesis of high performance $A_2$-$DA_1D$-$A_2$ type NFAs in organic photovoltaic applications. Moreover, it can be concluded that structural modulation of the core and the terminal units can be an effective strategy to tune the optoelectronic, charge transport, and overall photovoltaic properties of the non-fullerene acceptor molecules.

## Acknowledgement

The research was enabled with computational resources provided by the Indian Institute of Technology (Indian School of Mines), Dhanbad and Compute/Calcul Canada (computecanada.ca). L. Bhattacharya and S. Sahu is thankful to Indian Institute of Technology (Indian School of Mines), Dhanbad for research support. L. Bhattacharya is grateful to the Science and Engineering Board (SERB) India and University of Alberta, Canada for SERB-University of Alberta Overseas Visiting Doctoral Fellowship. L. Bhattacharya is also thankful to R. Khatua and S. Debata for discussions on crystal structure prediction and anisotropic mobility calculations.

## Supporting Information Available

The supporting information contains crystal structure prediction details, theory related to charge transfer integral, charge transfer rate and anisotropic mobility, Table S1-S8, and Figure S1-S14.

# Supporting Information

# Computational Design of Crescent Shaped Promising Non-Fullerene Acceptors with 2,3-quinoxaline,1,4-dihydro Core and Different Electron-withdrawing Terminal Units for Photovoltaic Applications

Labanya Bhattacharya,[1] Alex Brown[2], Sagar Sharma,[3] and Sridhar Sahu[1*]

[1]*Computational Materials Research Lab, Department of Physics, Indian Institute of Technology (Indian School of Mines), Dhanbad, Jharkhand-826004, India*

[2]*Department of Chemistry, University of Alberta, Edmonton, Alberta, T6G 2G2, Canada*

[3]*Department of Chemistry, S. B. Deorah College, Bora Service, Ulubari, Guwahati-781007, Assam, India*

E-mail: *sridharsahu@iitism.ac.in

**CONTENTS**

**Theory**

**Section S.1.** Crystal structure prediction

**Section S.2.** Theory related to charge transfer integral, charge transfer rate and anisotropic mobility

**Section S.3.** References

**List of Tables**

**Table S1.** The HOMO energy levels in eV of AQx-2 in the gas phase as determined using several functionals with the 6-31G(d,p) and 6-311G(d,p) basis sets

**Table S2.** Computed HOMO, LUMO energies and optical band gap ($E_{g,opt}$) of PM6 polymer donor at B3LYP-D3BJ/6-31G(d,p) theoretical level in the gas phase

**Table S3.** The maximum optical absorption wavelength ($\lambda_{abs}$) of AQx-2 in Chloroform solvent medium using several functionals with the 6-31G(d,p) basis set. All the computations at the B3LYP-D3BJ/6-31G(d,p) optimized geometry

**Table S4**. Dihedral angles (α,β,γ) in degrees and bond lengths ($d_1$, $d_2$, $d_3$, $d_4$) in Å of the designed NFAs as determined at the B3LYP-D3BJ/6-31G(d,p) level of theory in the gas phase

**Table S5.** Charge injection barriers in eV of the designed NFAs w.r.t Al electrode

**Table S6.** Computed electronic transition energy E (eV), absorption wavelength $\lambda_{abs}$ (nm), major configurations, oscillator strength (*f*), and light harvesting efficiency ($\eta_\lambda$) of the designed NFAs as determined at the B3LYP-D3BJ/6-31G(d,p) level of theory in the solvent (Chloroform) phase

**Table S7.** Space group and simulated unit cell parameters (a, b, c in Å and α, β, γ in deg) of the studied NFAs.

**Table S8.** Intermolecular charge transfer (CT) states, excitation energy (E), absorption wavelength ($\lambda_{abs}$), oscillator strength (*f*), Δr index, contribution rate of the main NTO pairs for the investigated PM6 donor-NFA blends or complexes at the CAM-B3LYP/6-31G(d,p) level of theory in solvent (Chloroform) phase

**List of Figures**

**Figure S1.** HOMO, LUMO energies and optical band gap ($E_{g,opt}$) of PM6 polymer donor at B3LYP-D3BJ/6-31G(d,p) theoretical level in the gas phase.

**Figure S2.** Molecular electrostatic potential (MEP) maps of the designed NFAs.

**Figure S3.** The contour plots of FMOs of the designed NFAs (AQx-2, **AQx-2-c**, **AQx-2-ct1**, **AQx-2-ct2**, and **AQx-2-ct3**; isosurface value= 0.02 a.u.) at the B3LYP-D3BJ/6-31G(d,p) level of theory.

**Figure S4.** The contour plots of FMOs of the designed NFAs (**AQx-2-ct4**, **AQx-2-ct5**, **AQx-2-ct6**, **AQx-2-ct7** and **AQx-2-ct8**; isosurface value= 0.02 a.u.) at the B3LYP-D3BJ/6-31G(d,p) level of theory.

**Figure S5.** Simulated charge density difference (Δρ) plots associated with the $S_0 \rightarrow S_1$ transition of the studied NFAs at the CAM-B3LYP/6-31G(d,p) level of theory in Chloroform solvent ( isosurface value 0.0002 a.u.). The green and blue regions correspond to positive and negative regions, respectively

**Figure S6.** Extending zones for the centroids C+/C- associated with the $S_0 \rightarrow S_1$ transition of the studied NFAs at the CAM-B3LYP/6-31G(d,p) level of theory in Chloroform solvent ( isosurface value 0.0002 a.u.). The green and blue regions correspond to positive and negative regions, respectively

**Figure S7.** Simulated transition density matrix (TDM) at the CAM-B3LYP/6-31G(d,p) level of theory associated with the $S_0 \rightarrow S_1$ transition of the studied NFAs (AQx-2, **AQx-2-ct1**, **AQx-2-ct2**, **AQx-2-ct3**, **AQx-2-ct4**, and **AQx-2-ct5**) in Chloroform solvent (the hydrogen atoms of all molecular systems are omitted) and the color bars are given on the right

**Figure S8.** Simulated transition density matrix (TDM) at the CAM-B3LYP/6-31G(d,p) level of theory associated with the $S_0 \rightarrow S_1$ transition of the studied NFAs (**AQx-2-ct6**, **AQx-2-ct7**, and **AQx-2-ct8**) in Chloroform solvent (the hydrogen atoms of all molecular systems are omitted) and the color bars are given on the right

**Figure S9.** The atom numbering of the designed NFAs (NFA:**AQx-2-c**) corresponding to TDM mapping. The cyan, yellow, blue, red and greenish yellow respresents carbon, sulfur, nitrogen, oxygen and fluorine atoms, respectively.

**Figure S10.** Simulated crystal structures showing different hopping channels and angle resolved anisotropic mobility of the studied NFAs (**AQx-2-ct1**, **AQx-2-ct2**)

**Figure S11.** Simulated crystal structures showing different hopping channels and angle resolved anisotropic mobility of the studied NFAs (**AQx-2-ct4**, and **AQx-2-ct5**)

**Figure S12.** Simulated crystal structures showing different hopping channels and angle resolved anisotropic mobility of the studied NFAs (**AQx-2-ct6**, **AQx-2-ct7**, and **AQx-2-ct8**)

**Figure S13.** The NTO analysis of the CT excited states of the investigated PM6/NFA (**AQx-2-ct1**, **AQx-2-ct2**, **AQx-2-ct4**, **AQx-2-ct5**) blends. The blue and red colors represent the positive and negative isosurfaces, respectively

**Figure S14.** The NTO analysis of the CT excited states of the investigated PM6/NFA (**AQx-2-ct6**, **AQx-2-ct7**, **AQx-2-ct8**) blends. The blue and red colors represent the positive and negative isosurfaces, respectively

**Theory**

**S.1. Crystal Structure Prediction**

Crystal structures of the designed NFAs have been predicted employing the most reliable Dreiding force field within the Polymorph Predictor module of the BIOVIA Materials Studio17 based on their optimized gas-phase conformations. During computation, the Polymorph Predictor quality is set to default fine settings, which anneals the sample in Monte Carlo simulation algorithm within a temperature ranges from 300.0 K to 100000.0 K, and a heating factor of 0.025. The maximum no of steps is 7000 and before cooling, 12 successive steps are accepted. All possible crystal structures of the studied NFAs are selected by energy minimization of crystals with the most common space groups such as $P2_1/C$, $P1$, $P\bar{1}$, $Pbca$, $P2_12_12_1$, $P2_1$, $C2/c$, $Pna2_1$, $Cc$, $Pbcn$, and $C2$ as registered in Cambridge Structural Database (CSD) [1, 2, 3]. For every NFA, the structure with the lowest lattice energy has been selected for further calculations.

**S.2. Theory related to charge transfer integral, charge transfer rate, and anisotropic mobility**

**S.2.1. Charge transfer rate:**

Based on Marcus Theory, the electron and hole transport in an organic semiconductor material can be described by the hoping mechanism. For the self-exchange transfer reaction where the neighboring molecules are equivalent, the free energy difference (ΔG) is nearly equal to zero. The rate of charge transfer (K) rate for electron (e) and hole (h) transport can be represented as [4, 5, 6],

$$k_{h/e} = \frac{2\pi V^2}{h} \left(\frac{\pi}{\lambda k_B T}\right)^{1/2} \exp\left(-\frac{\lambda}{4 k_B T}\right) \quad (1)$$

In the above equation, V is the charge transfer integral, λ is the reorganization energy, $k_B$ is the Boltzmann constant, and T is the temperature (T=300 K).

**S.2.2 Charge transfer integral:**

Charge transfer integrals are calculated based on the direct coupling (DC) method. The charge transfer integrals estimated using this method can provide more accurate results as this method takes into account the nonzero spatial overlap between the molecular orbitals of neighbor molecules [7, 8]. In the direct coupling method, the hole transfer integral or electron transfer integral is computed as,

$$V_{h/e} = \frac{J - S(e_1 + e_2)/2}{1 - S^2} \quad (2)$$

where J is the charge transfer integral, S is the overlap integral, and $e_1$, $e_2$, are the site energies of fragment 1 and 2, of the molecular dimer, respectively. The electronic coupling terms J, S, $e_1$, and $e_2$ can be defined as;

$$J = \langle \varphi_1 | H_{KS} | \varphi_2 \rangle \quad (3)$$

$$S_{12} = \langle \varphi_1 | \varphi_2 \rangle \quad (4)$$

$$e_1 = \langle \varphi_1 | H_{KS} | \varphi_1 \rangle \quad (5)$$

$$e_2 = \langle \varphi_2 | H_{KS} | \varphi_2 \rangle \quad (6)$$

In the electron transfer integral calculations, $\varphi_1$ is the LUMO of fragment 1, and $\varphi_2$ is the LUMO of fragment 2. Similarly, for the hole transport, $\varphi_1$ is the HOMO of fragment 1, and $\varphi_2$ is the HOMO of fragment 2.

### S.2.3 Reorganization energy:

The internal reorganization ($\lambda_i$) energies of all the investigated NFAs during hole, and electron transfer are evaluated using the adiabatic potential energy surface (APS) method. Using that method, the internal reorganization energy for the hole ($\lambda_h$) and electron ($\lambda_e$) transfer can then be described as [6];

$$\lambda_h = [E^+(M^0) - E^+(M^+)] + [E^0(M^+) - E^0(M^0)] \quad (7)$$

$$\lambda_e = [E^0(M^-) - E^0(M^0)] + [E^-(M^0) - E^-(M^-)] \quad (8)$$

The minimum energy of the neutral state is described as $E^0(M^0)$, and for cationic and anionic states, the minimum energies are $E^+(M^+)$ and $E^-(M^-)$, respectively. During the charge transfer process, geometry changes arise in four possible energy states such as $E^-(M^0)$, $E^+(M^0)$, $E^0(M^+)$, and $E^0(M^-)$. The $E^-(M^0)$ and $E^+(M^0)$ are the total energy of the anionic/cationic states with the geometries of the neutral species. The $E^0(M^+)$ and $E^0(M^-)$ define the total energy of the neutral state with respect to the geometry of cationic/anionic species.

### S.2.4 Anisotropic Mobility:

At high temperature, the charge transport in the disordered organic crystals is found to be anisotropic in nature, because of the increased fluctuations in the intermolecular interactions. The angular resolution anisotropic mobility ($\mu_\varphi$) in organic crystals is subject to the molecular packing in a particular plane. In accordance with the Einstein relation, the anisotropic mobility ($\mu_\varphi$) can be evaluated as [9, 10],

$$\mu_\varphi = \frac{e}{2k_BT} \sum_i k_i r_i^2 P_i Cos^2\gamma_i Cos^2(\theta_i - \varphi) \qquad (9)$$

In the above equation, $r_i$ is the intermolecular distance associated with i'th hopping pathway and $P_i$ is the hoping probability of the transport channel. $\gamma_i$ is the angle made by the corresponding hoping pathways with respect to the basal plane of the dimers. $\theta_i$ is the angle of the hoping path relative to the crytallographic axis, $\varphi$ is the angle of orientation of the conducting channel relative to the crystallographic axis and ($\theta_i$-$\varphi$) denotes the angle between the conducting channel and hoping pathway.

**Table S1.** The HOMO energy levels in eV of AQx-2 in the gas phase as determined using several functionals with the 6-31G(d,p) and 6-311G(d,p) basis sets.

| Method | Basis set | | Exp. |
|---|---|---|---|
| | 6-31G(d,p) | 6-311G(d,p) | |
| B3LYP | -5.51 | -5.75 | -5.62 |
| B3LYP-D3BJ | -5.52 | -5.76 | |
| PBE0 | -5.73 | -5.91 | |
| MPW91PW91 | -5.75 | -5.93 | |
| M06 | -5.80 | -6.01 | |
| ωB97XD | -7.10 | -7.29 | |

**Table S2.** Computed HOMO, LUMO energies and optical band gap ($E_{g,opt}$) of PM6 polymer donor at B3LYP-D3BJ/6-31G(d,p) theoretical level in the gas phase

| Degree of polym. (n) | HOMO (eV) | LUMO (eV) | $E_{g,opt}$ (eV) | LUMO (HOMO+ $E_{g,opt}$) (eV) |
|---|---|---|---|---|
| 1 | -5.15 | -2.46 | 2.33 | -2.82 |
| 2 | -5.02 | -2.66 | 2.01 | -3.01 |
| 3 | -4.98 | -2.72 | 1.90 | -3.08 |
| 4 | -4.96 | -2.75 | 1.87 | -3.09 |
| ∞ | -4.90 | -2.84 | 1.78 | -3.12 |

**Table S3.** The maximum optical absorption wavelength ($\lambda_{abs}$) of AQx-2 in Chloroform solvent medium using several functionals with the 6-31G(d,p) basis set. All the computations at the B3LYP-D3BJ/6-31G(d,p) optimized geometry

| Method | State | λ abs (nm) | Major. Conf. | Oscillator Strength (*f*) | Exp. data |
|---|---|---|---|---|---|
| B3LYP-D3BJ | $S_0 \rightarrow S_1$ | 716 | H → L (99%) | 2.2623 | 732 nm |
| PBE0 | $S_0 \rightarrow S_1$ | 684 | H → L (97%) | 2.3916 | |
| M06 | $S_0 \rightarrow S_1$ | 681 | H → L (95%) | 2.3614 | |
| LC-ωPBE | $S_0 \rightarrow S_1$ | 528 | H → L (66%) | 2.6689 | |
| LC-BLYP | $S_0 \rightarrow S_1$ | 526 | H → L (65%) | 2.6810 | |
| CAM-B3LYP | $S_0 \rightarrow S_1$ | 587 | H → L (80%) | 2.6644 | |
| ωB97XD | $S_0 \rightarrow S_1$ | 566 | H → L (73%) | 2.6738 | |

**Table S4.** Dihedral angles (α,β,γ) in degrees and bond lengths ($d_1,d_2,d_3,d_4$) in Å of the designed NFAs as determined at the B3LYP-D3BJ/6-31G(d,p) level of theory in the gas phase

| Comp. | α | β | γ | $d_1$ | $d_2$ | $d_3$ | $d_4$ |
|---|---|---|---|---|---|---|---|
| AQx-2 (Ref.) | 1.36 | 1.87 | 2.99 | 1.408 | 1.382 | 1.408 | 1.382 |
| **AQx-2-c** | 1.34 | 1.87 | 1.91 | 1.411 | 1.379 | 1.411 | 1.379 |
| **AQx-2-ct1** | 1.37 | 1.91 | 1.94 | 1.409 | 1.381 | 1.409 | 1.381 |
| **AQx-2-ct2** | 1.31 | 1.93 | 1.91 | 1.410 | 1.381 | 1.410 | 1.381 |
| **AQx-2-ct3** | 1.31 | 1.93 | 1.93 | 1.409 | 1.382 | 1.409 | 1.382 |
| **AQx-2-ct4** | 1.36 | 1.83 | 1.90 | 1.411 | 1.380 | 1.411 | 1.380 |
| **AQx-2-ct5** | 1.45 | 1.93 | 1.83 | 1.415 | 1.375 | 1.415 | 1.375 |
| **AQx-2-ct6** | 1.32 | 1.90 | 1.68 | 1.424 | 1.365 | 1.424 | 1.365 |
| **AQx-2-ct7** | 1.40 | 1.96 | 1.73 | 1.421 | 1.366 | 1.421 | 1.366 |
| **AQx-2-ct8** | 2.67 | 3.20 | 1.87 | 1.417 | 1.375 | 1.417 | 1.375 |

**Table S5.** Charge injection barriers in eV of the designed NFAs w.r.t Al electrode

| Comp. | Hole injection barrier | Electron injection barrier |
|---|---|---|
| AQx-2 (Ref.) | 1.22 | 0.63 |
| **AQx-2-c** | 1.41 | 0.41 |
| **AQx-2-ct1** | 1.46 | 0.32 |
| **AQx-2-ct2** | 1.35 | 0.44 |
| **AQx-2-ct3** | 1.41 | 0.35 |
| **AQx-2-ct4** | 1.33 | 0.49 |
| **AQx-2-ct5** | 1.25 | 0.68 |
| **AQx-2-ct6** | 0.98 | 1.11 |
| **AQx-2-ct7** | 1.24 | 0.80 |
| **AQx-2-ct8** | 1.47 | 0.64 |

**Table S6.** Computed electronic transition energy E (eV), absorption wavelength $\lambda_{abs}$ (nm), major configurations, oscillator strength ($f$), and light harvesting efficiency ($\eta_\lambda$) of the designed NFAs as determined at the B3LYP-D3BJ/6-31G(d,p) level of theory in the solvent (Chloroform) phase

| Compound | State | E (eV) | $\lambda_{abs}$ (nm) | major config. | $f$ | $\eta_\lambda$ |
|---|---|---|---|---|---|---|
| AQx-2 (ref) | $S_0 \rightarrow S_1$ | 1.73 | 716 | H → L (99%) | 2.2623 | 0.9945 |
|  | $S_0 \rightarrow S_7$ | 2.55 | 486 | H-2 → L (75%) | 0.5170 |  |
| **AQx-2-c** | $S_0 \rightarrow S_1$ | 1.77 | 701 | H → L (98%) | 2.0966 | 0.9920 |
|  | $S_0 \rightarrow S_8$ | 2.58 | 480 | H-2 → L (52%) | 0.3434 |  |
| **AQx-2-ct1** | $S_0 \rightarrow S_1$ | 1.74 | 713 | H → L (98%) | 2.1140 | 0.9923 |
|  | $S_0 \rightarrow S_9$ | 2.54 | 487 | H-2 → L (84%) | 0.5382 |  |
| **AQx-2-ct2** | $S_0 \rightarrow S_1$ | 1.72 | 719 | H → L (98%) | 2.2204 | 0.9940 |
|  | $S_0 \rightarrow S_5$ | 2.49 | 497 | H-2 → L (97%) | 0.3917 |  |
| **AQx-2-ct3** | $S_0 \rightarrow S_1$ | 1.70 | 730 | H → L (98%) | 2.2345 | 0.9942 |
|  | $S_0 \rightarrow S_5$ | 2.47 | 502 | H-2 → L (97%) | 0.4211 |  |
| **AQx-2-ct4** | $S_0 \rightarrow S_1$ | 1.77 | 700 | H → L (98%) | 2.2034 | 0.9937 |
|  | $S_0 \rightarrow S_5$ | 2.55 | 486 | H-2 → L (98%) | 0.3830 |  |
| **AQx-2-ct5** | $S_0 \rightarrow S_1$ | 1.84 | 673 | H → L (98%) | 1.7130 | 0.9806 |
|  | $S_0 \rightarrow S_5$ | 2.18 | 568 | H → L+2 (95%) | 0.5074 |  |
| **AQx-2-ct6** | $S_0 \rightarrow S_1$ | 1.99 | 622 | H → L (99%) | 1.7252 | 0.9812 |
|  | $S_0 \rightarrow S_5$ | 2.71 | 457 | H-2 → L (97%) | 0.8332 |  |
| **AQx-2-ct7** | $S_0 \rightarrow S_1$ | 1.96 | 633 | H → L (99%) | 1.8702 | 0.9865 |
|  | $S_0 \rightarrow S_5$ | 2.68 | 463 | H-2 → L (98%) | 0.8750 |  |
| **AQx-2-ct8** | $S_0 \rightarrow S_1$ | 2.08 | 597 | H-1 → L (98%) | 0.0407 |  |
|  | $S_0 \rightarrow S_2$ | 2.14 | 579 | H → L (97%) | 1.6399 | 0.9771 |
|  | $S_0 \rightarrow S_5$ | 3.07 | 404 | H-2 → L (90%) | 0.6101 |  |

**Table S7.** Space group and simulated unit cell parameters (a, b, c in Å and α, β, γ in deg) of the studied NFAs.

| Compound | Space group | Lattice parameters | | | | | |
|---|---|---|---|---|---|---|---|
| | | a | b | c | α | β | γ |
| AQx-2 (Ref.) | P2$_1$ | 19.86 | 18.68 | 12.77 | 90.00 | 109.00 | 90.00 |
| **AQx-2-c** | P1 | 13.29 | 12.60 | 14.77 | 114.00 | 90.00 | 85.00 |
| **AQx-2-ct1** | P1 | 13.74 | 15.03 | 12.48 | 66.54 | 79.89 | 90.24 |
| **AQx-2-ct2** | P1 | 8.67 | 15.32 | 24.83 | 101.56 | 96.99 | 121.25 |
| **AQx-2-ct3** | P1 | 13.54 | 16.21 | 14.68 | 124.36 | 108.21 | 86.12 |
| **AQx-2-ct4** | P1 | 21.38 | 13.21 | 11.92 | 107.09 | 122.64 | 107.13 |
| **AQx-2-ct5** | Pna2$_1$ | 25.40 | 19.21 | 18.08 | 90.00 | 90.00 | 90.00 |
| **AQx-2-ct6** | P2$_1$ | 13.97 | 17.53 | 18.29 | 90.00 | 111.55 | 90.00 |
| **AQx-2-ct7** | P$\bar{1}$ | 20.07 | 17.71 | 13.53 | 105.30 | 105.42 | 96.95 |
| **AQx-2-ct8** | P$\bar{1}$ | 13.45 | 28.09 | 10.31 | 73.35 | 83.03 | 87.70 |

**Table S8.** Intermolecular charge transfer (CT) states, excitation energy (E), absorption wavelength ($\lambda_{abs}$), oscillator strength ($f$), Δr index, contribution rate of the main NTO pairs for the investigated PM6 donor-NFA blends or complexes at the CAM-B3LYP/6-31G(d,p) level of theory in solvent (Chloroform) phase

| Complex | CT state | E (eV) | $\lambda_{abs}$ (nm) | $f$ | Δr (Å) | NTO contribution rate |
|---|---|---|---|---|---|---|
| PM6:AQx-2 | S28 | 3.82 | 325 | 0.0375 | 8.4166 | 706 → 707 (60.78%) |
| **PM6:AQx-2-c** | S38 | 4.10 | 302 | 0.0167 | 15.3715 | 714 → 715 (91.01%) |
| **PM6:AQx-2-ct1** | S30 | 3.77 | 329 | 0.0621 | 6.1961 | 730 → 731 (78.19%) |
| **PM6:AQx-2-ct2** | S29 | 3.75 | 331 | 0.0054 | 9.0764 | 740 → 741 (81.22%) |
| **PM6:AQx-2-ct3** | S33 | 3.88 | 319 | 0.0546 | 8.0211 | 756 → 757 (68.31%) |
| **PM6:AQx-2-ct4** | S30 | 3.86 | 321 | 0.0979 | 8.6461 | 700 → 701 (62.93%) |
| **PM6:AQx-2-ct5** | S38 | 4.02 | 308 | 0.0891 | 6.8833 | 700 → 701 (65.04%) |
| **PM6:AQx-2-ct6** | S23 | 3.88 | 320 | 0.0177 | 8.7614 | 682 → 683 (83.69%) |
| **PM6:AQx-2-ct7** | S19 | 3.80 | 326 | 0.0096 | 7.4638 | 698 → 699 (77.28%) |
| **PM6:AQx-2-ct8** | S30 | 4.27 | 290 | 0.0080 | 9.1777 | 632 → 633 (82.82%) |

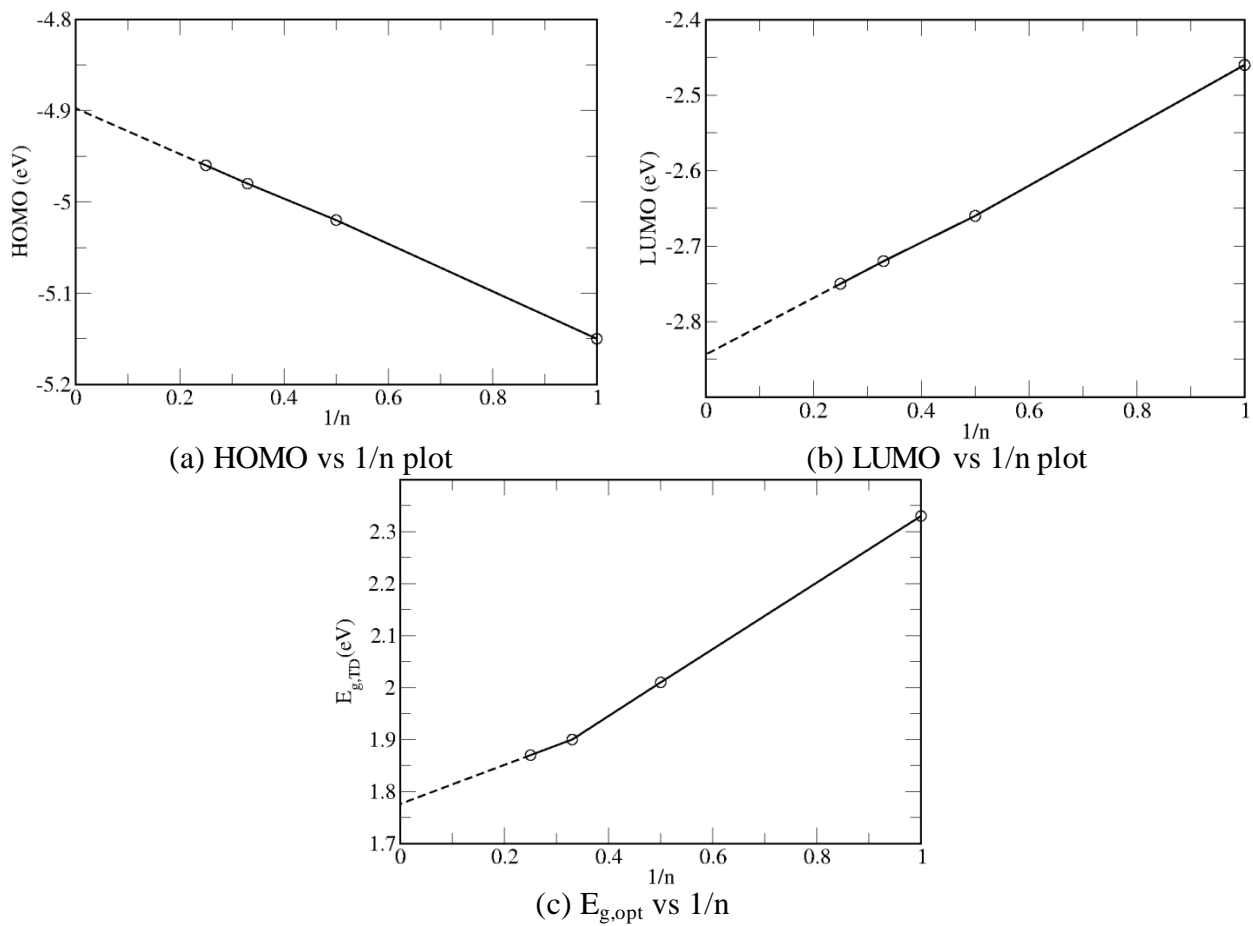

(a) HOMO vs 1/n plot

(b) LUMO vs 1/n plot

(c) $E_{g,opt}$ vs 1/n

**Figure S1.** HOMO, LUMO energies and optical band gap ($E_{g,opt}$) of PM6 polymer donor at B3LYP-D3BJ/6-31G(d,p) theoretical level in the gas phase

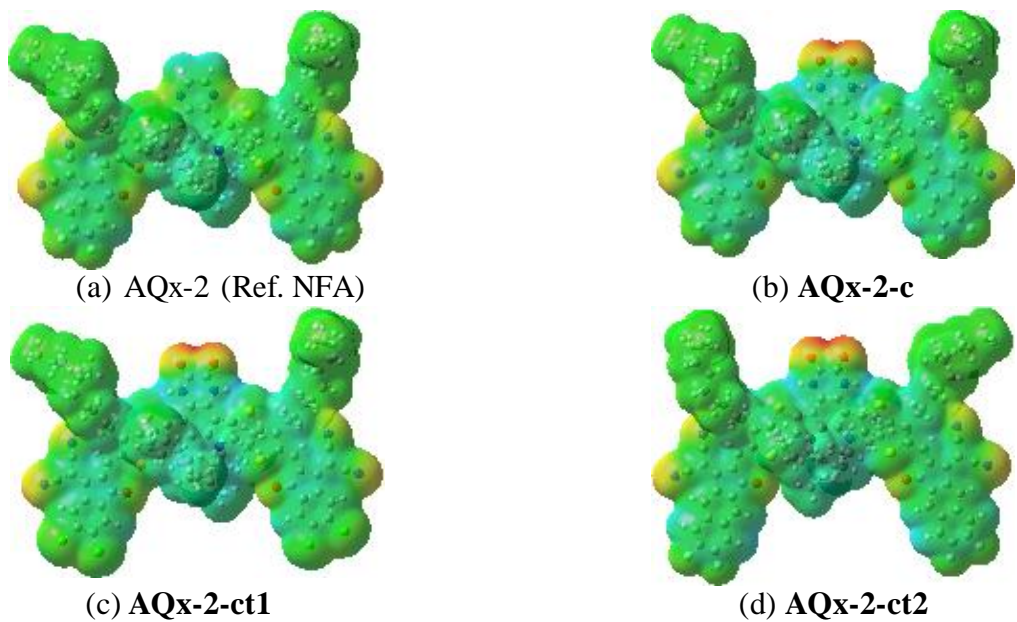

(a) AQx-2 (Ref. NFA)

(b) **AQx-2-c**

(c) **AQx-2-ct1**

(d) **AQx-2-ct2**

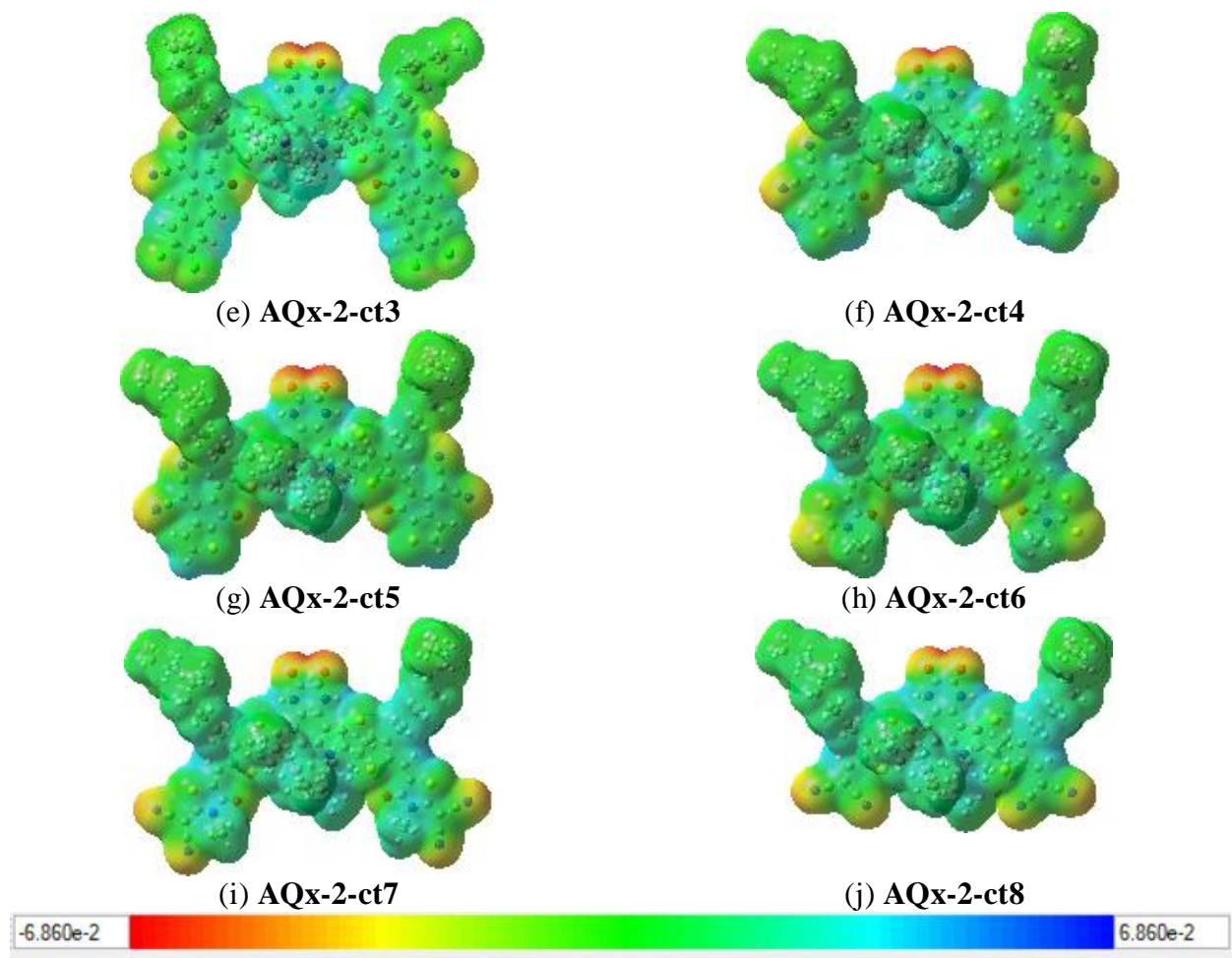

(e) **AQx-2-ct3**  (f) **AQx-2-ct4**

(g) **AQx-2-ct5**  (h) **AQx-2-ct6**

(i) **AQx-2-ct7**  (j) **AQx-2-ct8**

**Figure S2.** Molecular electrostatic potential (MEP) maps of the designed NFAs.

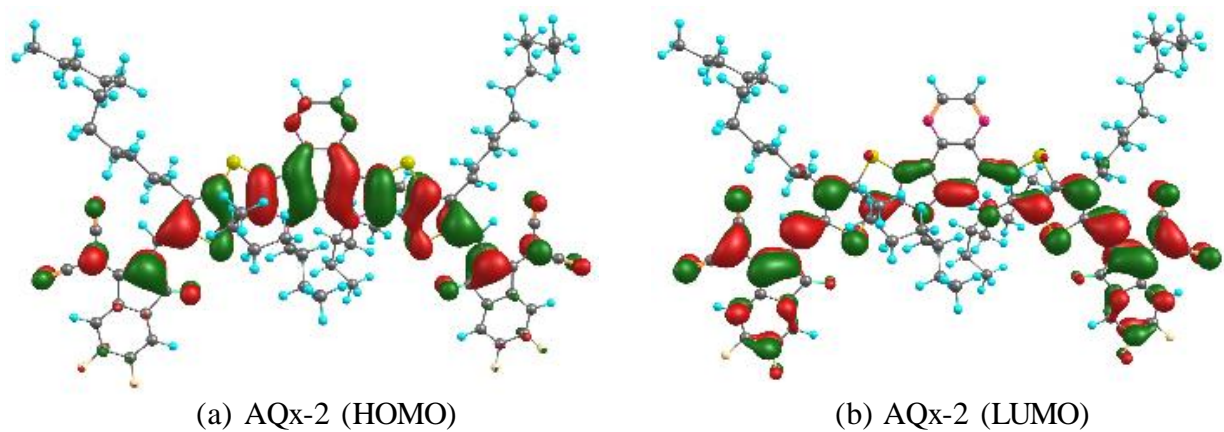

(a) AQx-2 (HOMO)  (b) AQx-2 (LUMO)

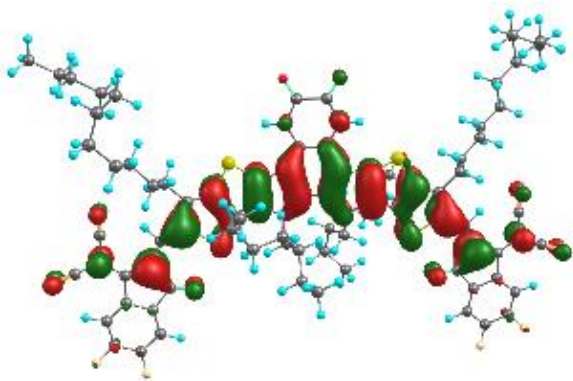
(C) **AQx-2-c** (HOMO)

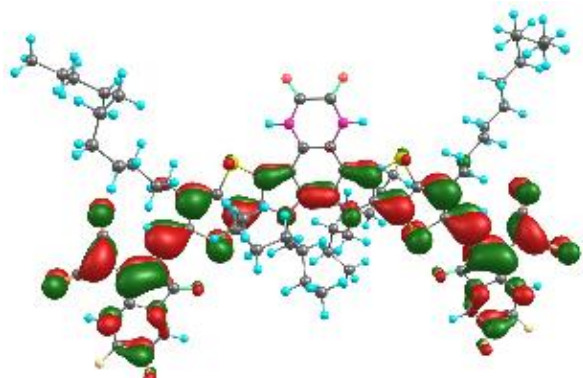
(d) **AQx-2-c** (LUMO)

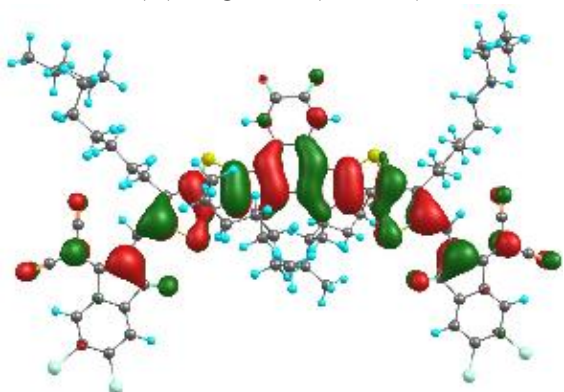
(e) **AQx-2-ct1** (HOMO)

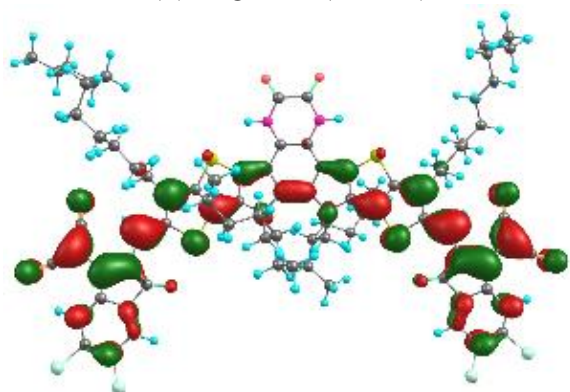
(f) **AQx-2-ct1** (LUMO)

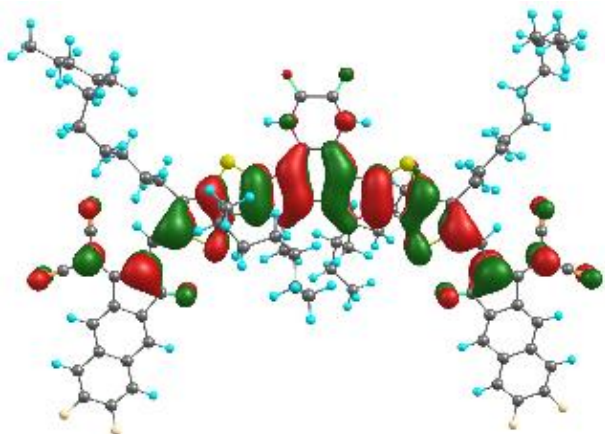
(g) **AQx-2-ct2** (HOMO)

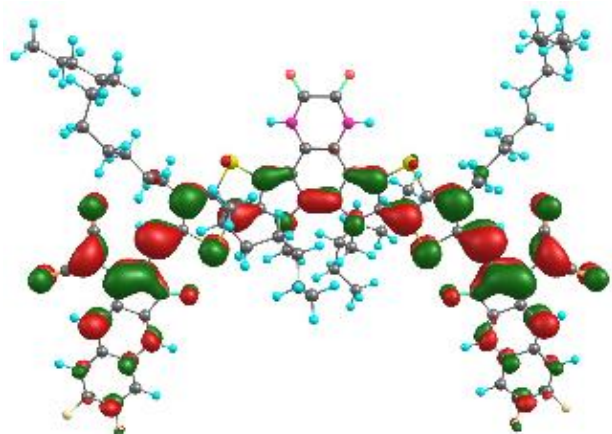
(h) **AQx-2-ct2** (LUMO)

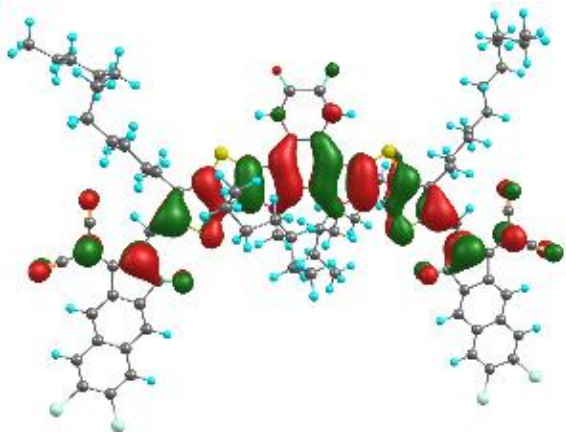 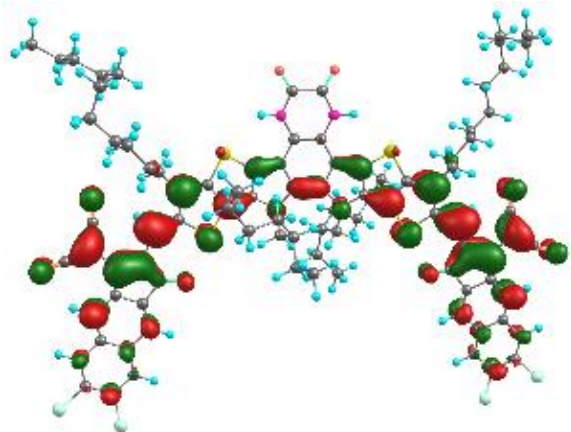

(i) **AQx-2-ct3** (HOMO)  (j) **AQx-2-ct3** (LUMO)

**Figure S3.** The contour plots of FMOs of the designed NFAs (AQx-2, **AQx-2-c**, **AQx-2-ct1**, **AQx-2-ct2**, and **AQx-2-ct3**; isosurface value= 0.02 a.u.) at the B3LYP-D3BJ/6-31G(d,p) level of theory.

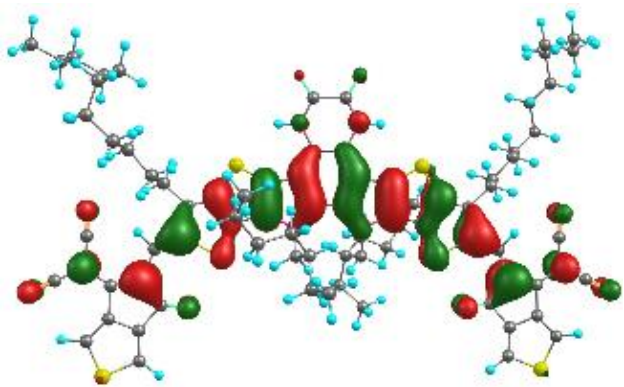 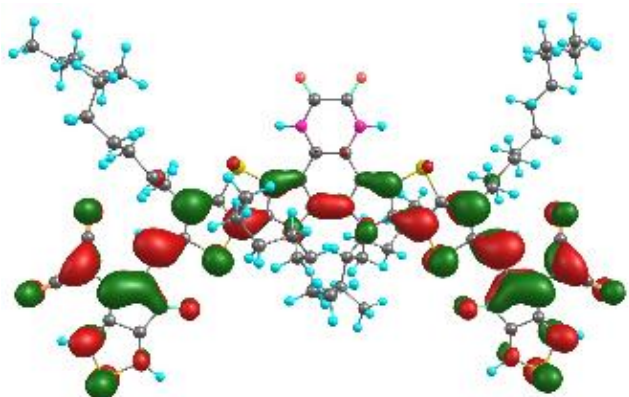

(a) **AQx-2-ct4** (HOMO)  (b) **AQx-2-ct4** (LUMO)

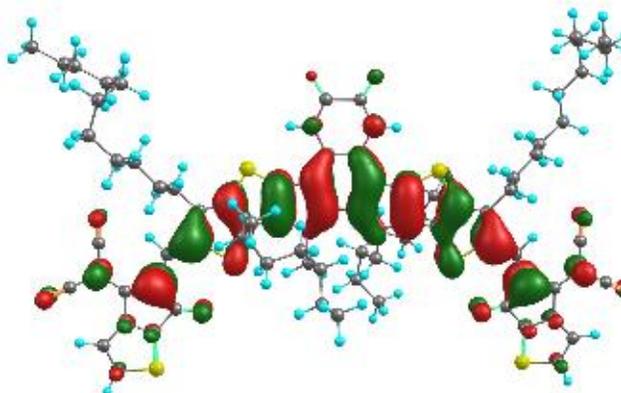 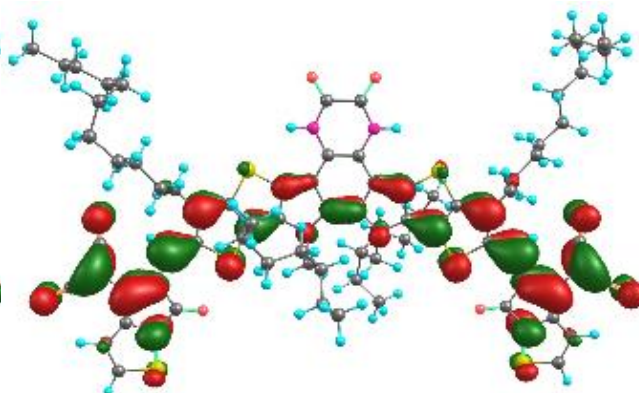

(c) **AQx-2-ct5** (HOMO)  (d) **AQx-2-ct5** (LUMO)

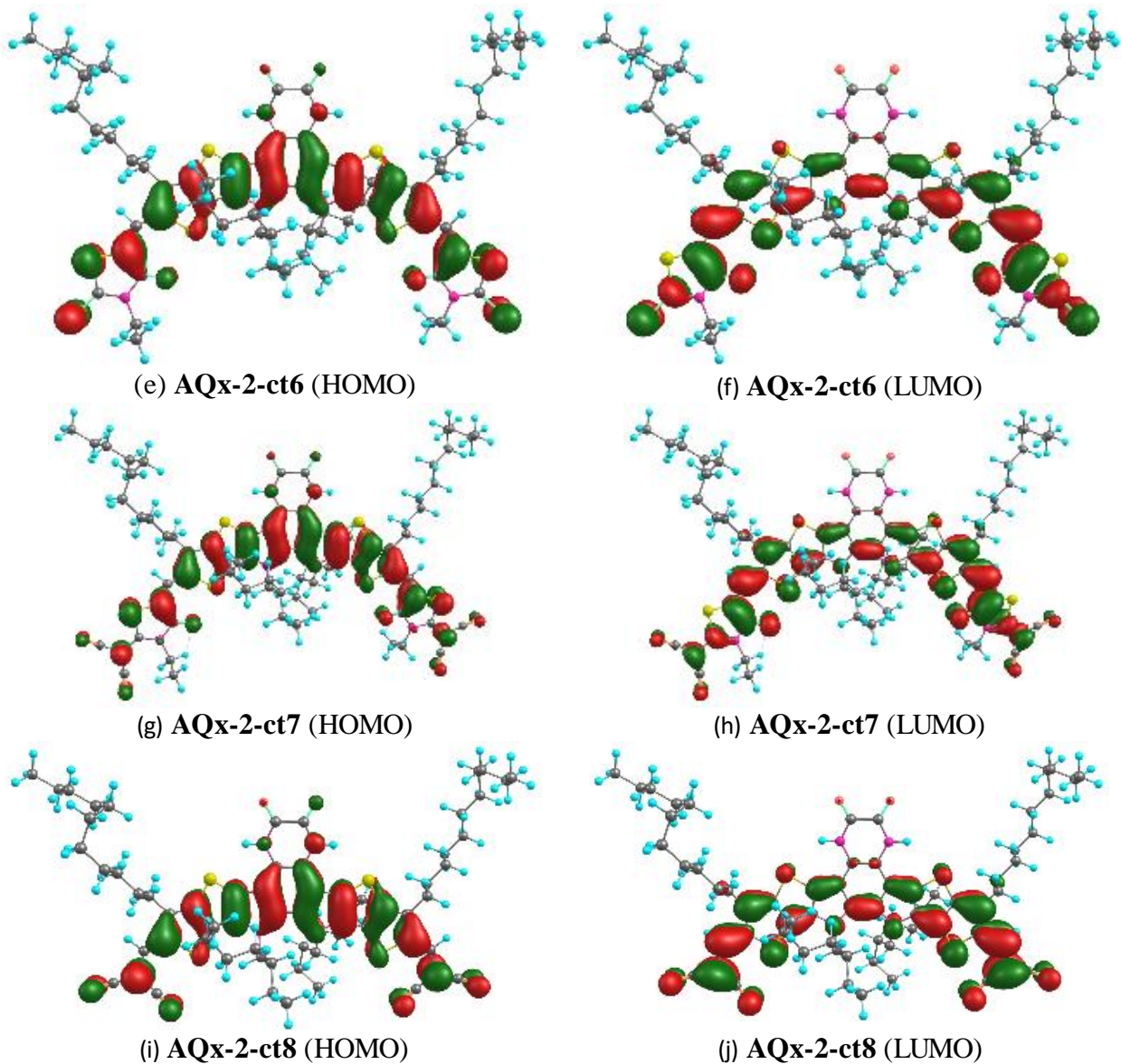

(e) **AQx-2-ct6** (HOMO)      (f) **AQx-2-ct6** (LUMO)

(g) **AQx-2-ct7** (HOMO)      (h) **AQx-2-ct7** (LUMO)

(i) **AQx-2-ct8** (HOMO)      (j) **AQx-2-ct8** (LUMO)

**Figure S4.** The contour plots of FMOs of the designed NFAs (**AQx-2-ct4**, **AQx-2-ct5**, **AQx-2-ct6**, **AQx-2-ct7** and **AQx-2-ct8**; isosurface value= 0.02 a.u.) at the B3LYP-D3BJ/6-31G(d,p) level of theory

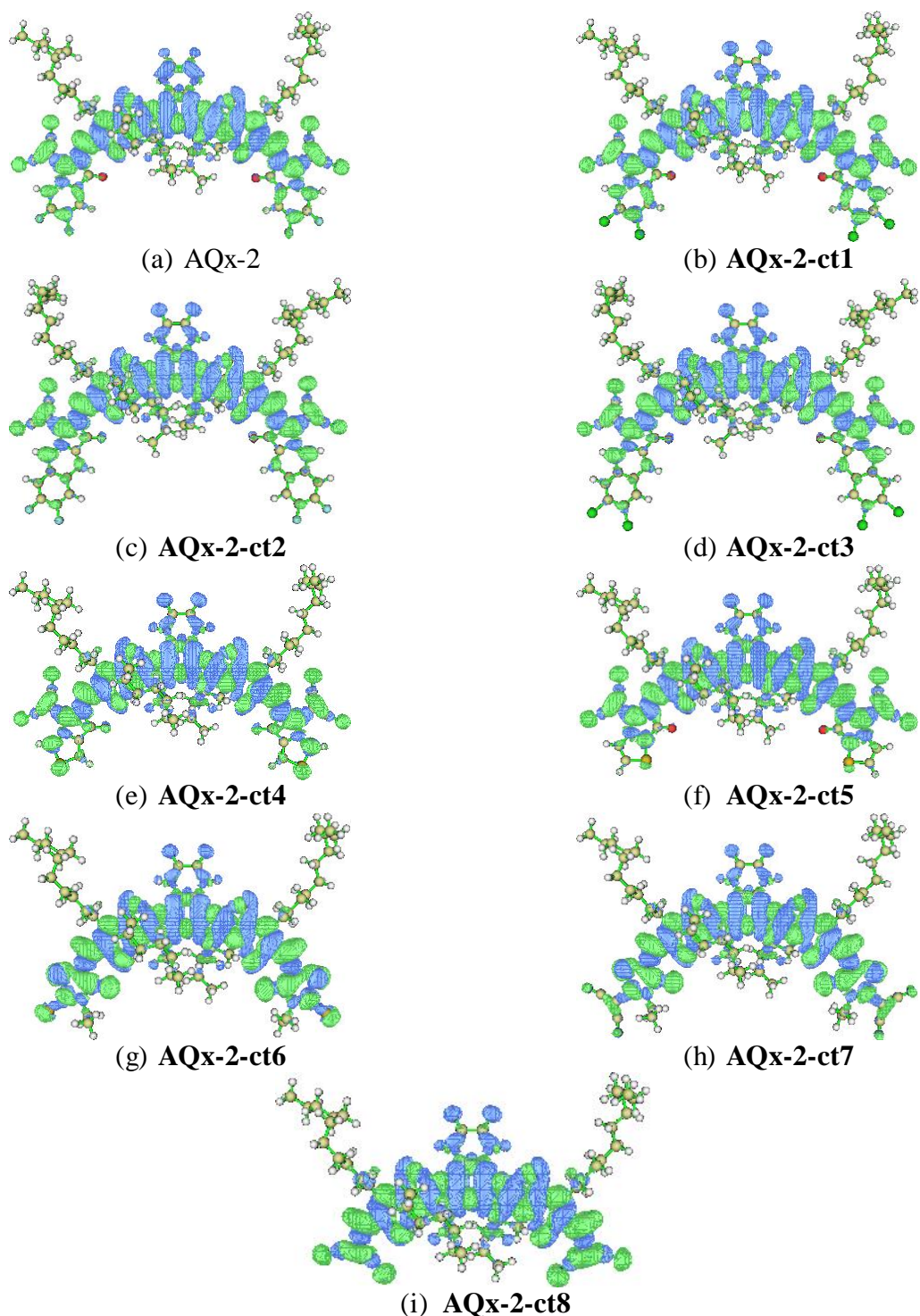

**Figure S5**. Simulated charge density difference (Δρ) plots associated with the S$_0$→S$_1$ transition of the studied NFAs at the CAM-B3LYP/6-31G(d,p) level of theory in Chloroform solvent ( isosurface value 0.0002 a.u.). The green and blue regions correspond to positive and negative regions, respectively

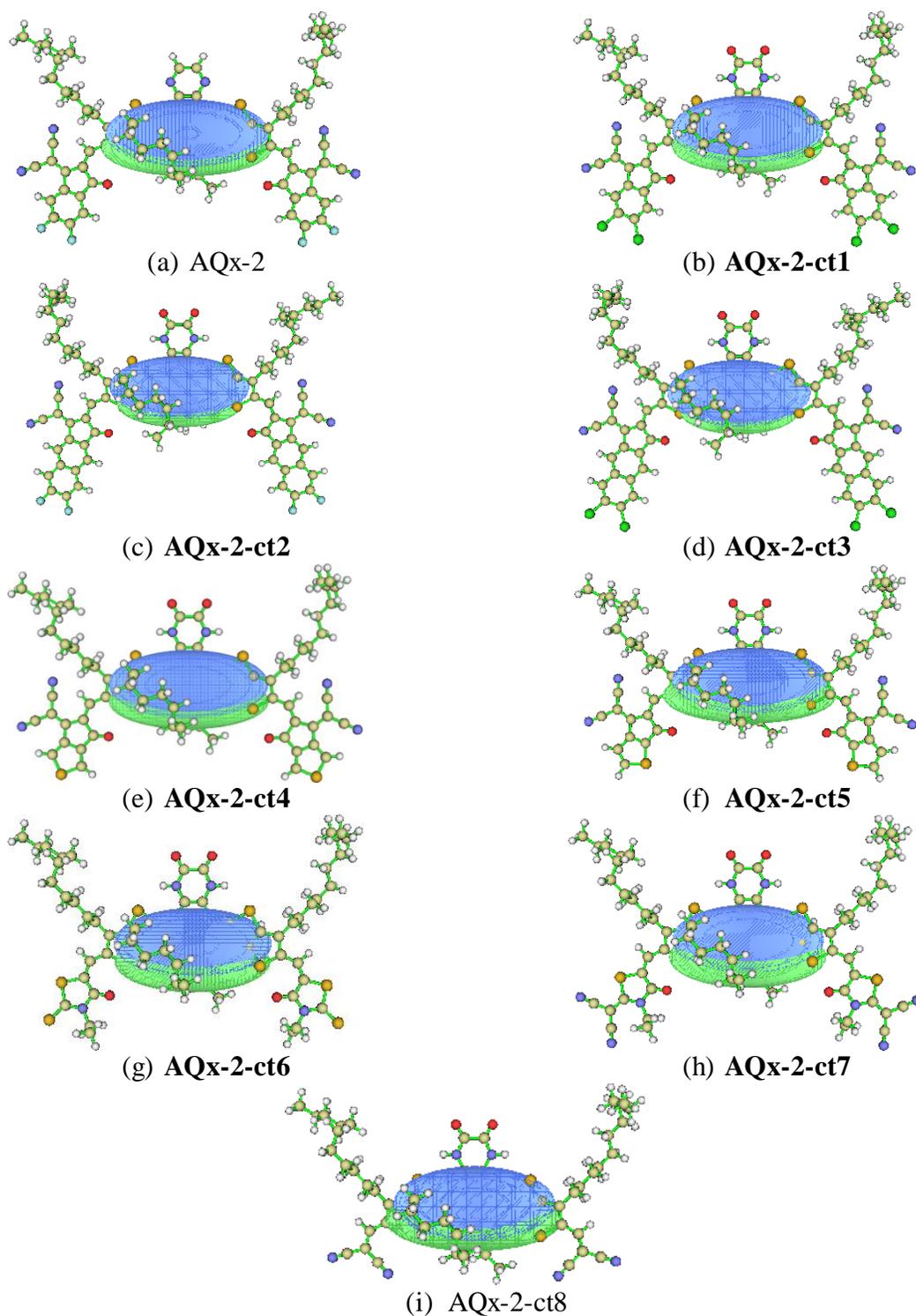

**Figure S6.** Extending zones for the centroids C+/C- associated with the $S_0{\rightarrow}S_1$ transition of the studied NFAs at the CAM-B3LYP/6-31G(d,p) level of theory in Chloroform solvent (isosurface value 0.0002 a.u.). The green and blue regions correspond to positive and negative regions, respectively

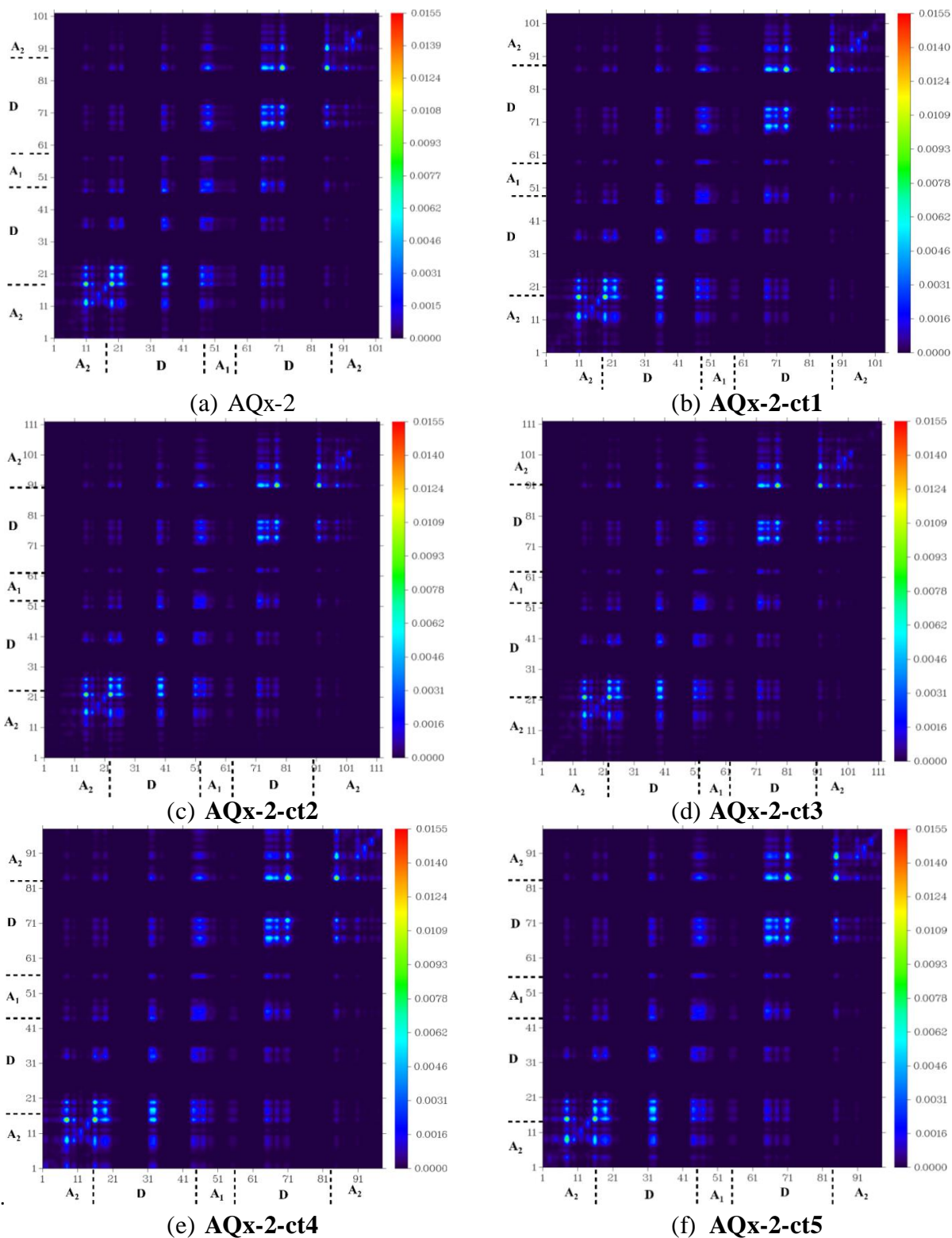

(a) AQx-2  (b) **AQx-2-ct1**
(c) **AQx-2-ct2**  (d) **AQx-2-ct3**
(e) **AQx-2-ct4**  (f) **AQx-2-ct5**

**Figure S7.** Simulated transition density matrix (TDM) at the CAM-B3LYP/6-31G(d,p) level of theory associated with the $S_0 \rightarrow S_1$ transition of the studied NFAs (AQx-2, **AQx-2-ct1**, **AQx-2-ct2**, **AQx-2-ct3**, **AQx-2-ct4**, and **AQx-2-ct5**) in Chloroform solvent (the hydrogen atoms of all molecular systems are

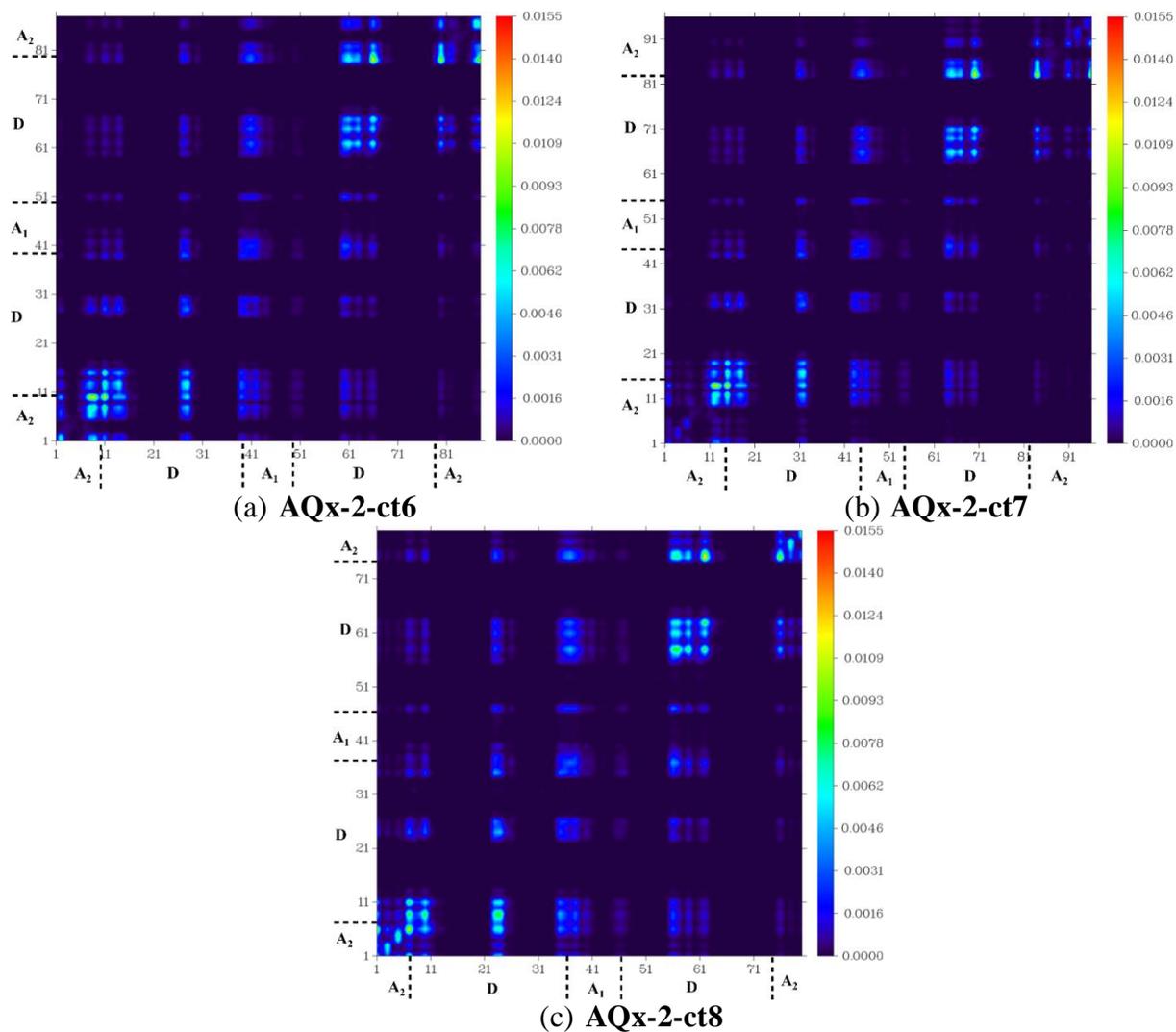

**Figure S8.** Simulated transition density matrix (TDM) at the CAM-B3LYP/6-31G(d,p) level of theory associated with the $S_0 \rightarrow S_1$ transition of the studied NFAs (**AQx-2-ct6**, **AQx-2-ct7**, and **AQx-2-ct8**) in Chloroform solvent (the hydrogen atoms of all molecular systems are omitted) and the color bars are given on the right

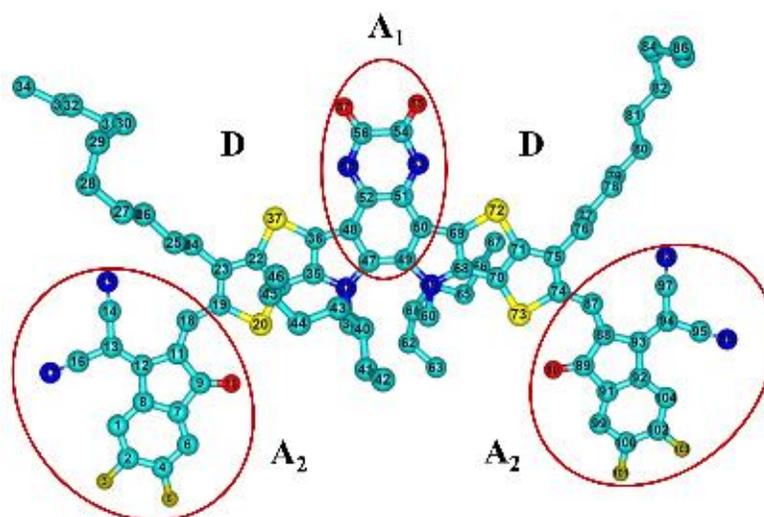

**Figure S9.** The atom numbering of the designed NFAs (NFA: **AQx-2-c**) corresponding to TDM mapping. The cyan, yellow, blue, red and greenish yellow represent carbon, sulfur, nitrogen, oxygen and fluorine atoms, respectively.

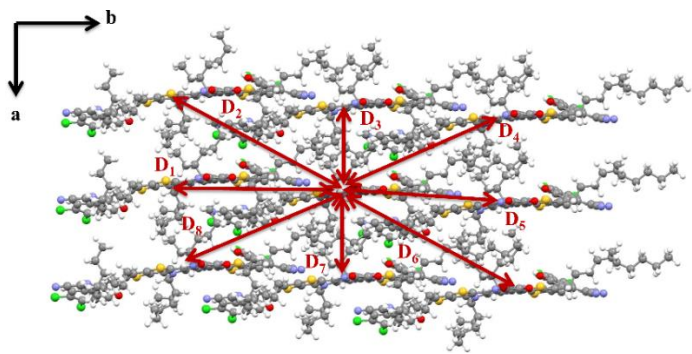
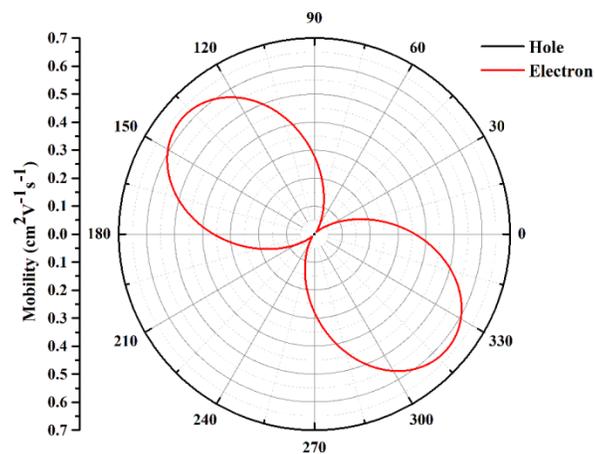

(a) **AQx-2-ct1 (packing)**  (b) **AQx-2-ct1 (mob)**

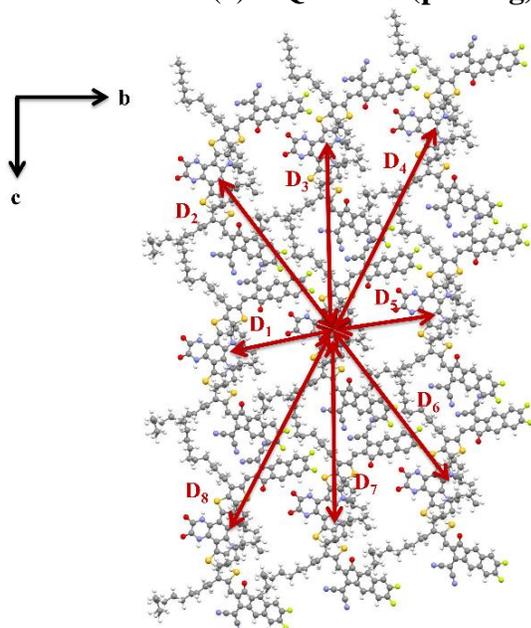
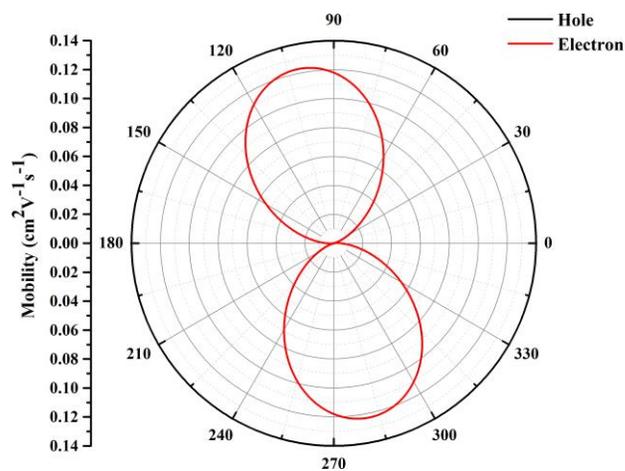

(c) **AQx-2-ct2 (packing)**  (d) **AQx-2-ct2 (mob)**

**Figure S10.** Simulated crystal structures showing different hopping channels and angle resolved anisotropic mobility of the studied NFAs (**AQx-2-ct1**, **AQx-2-ct2**)

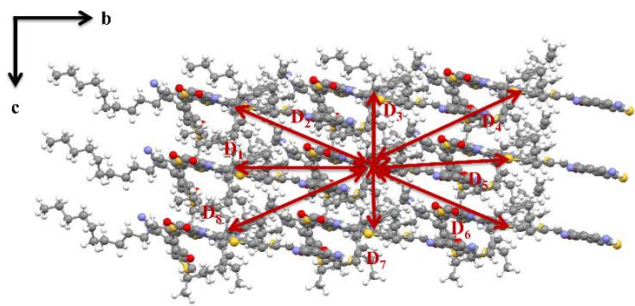
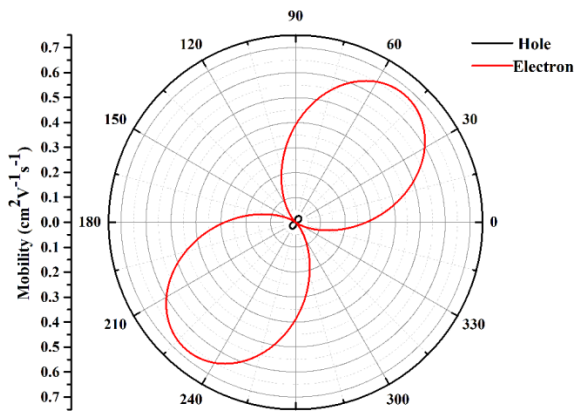

(a) **AQx-2-ct4 (packing)**　　　　　(b) **AQx-2-ct4 (mob)**

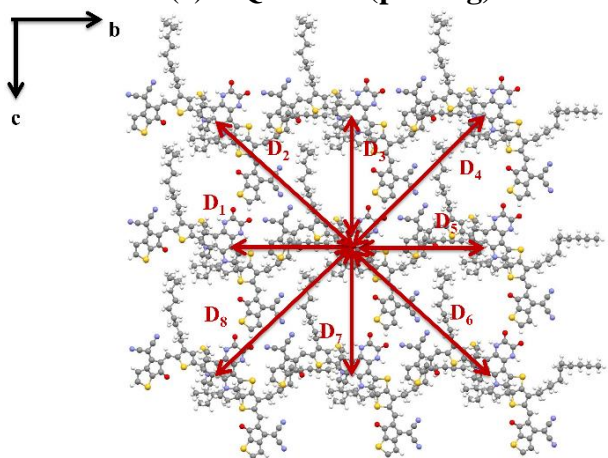
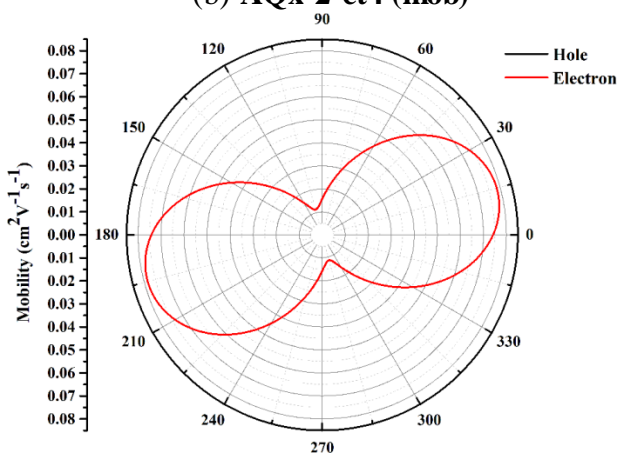

(c) **AQx-2-ct5 (packing)**　　　　　(d) **AQx-2-ct5 (mob)**

**Figure S11.** Simulated crystal structures showing different hopping channels and angle resolved anisotropic mobility of the studied NFAs (**AQx-2-ct4**, and **AQx-2-ct5**)

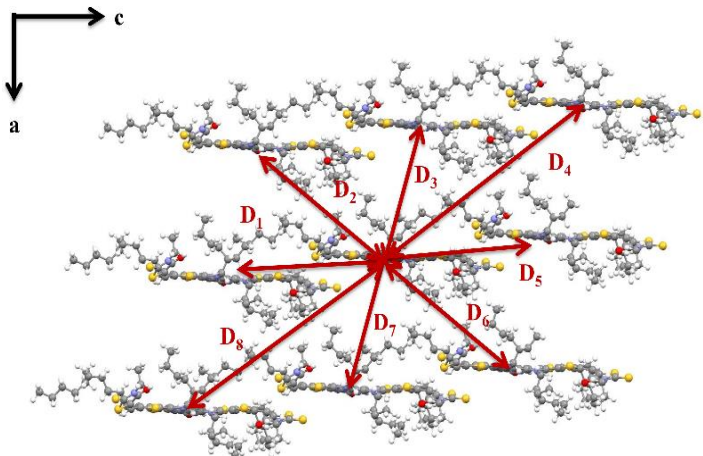
(a) **AQx-2-ct6 (packing)**

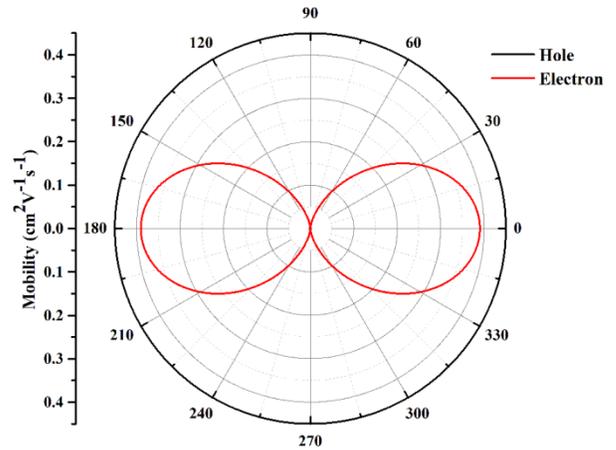
(b) **AQx-2-ct6 (mob)**

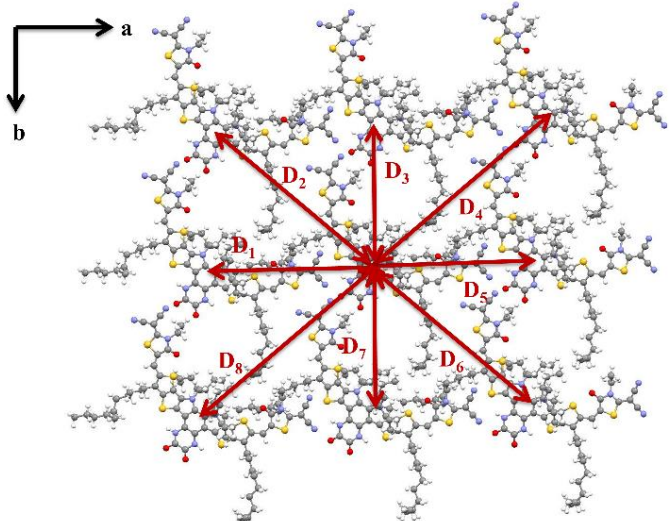
(c) **AQx-2-ct7 (packing)**

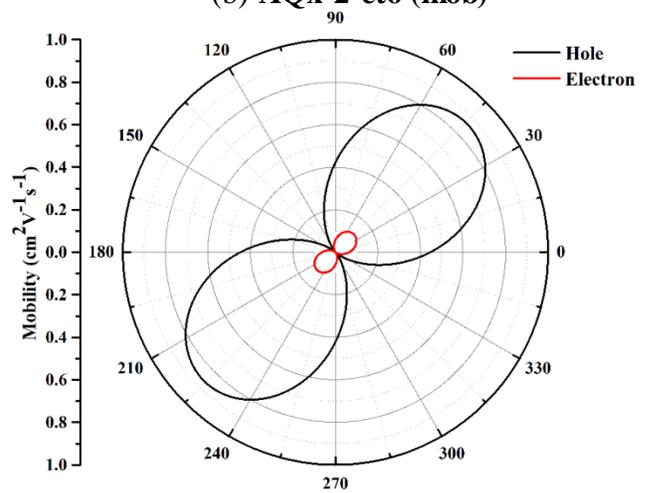
(d) **AQx-2-ct7 (mob)**

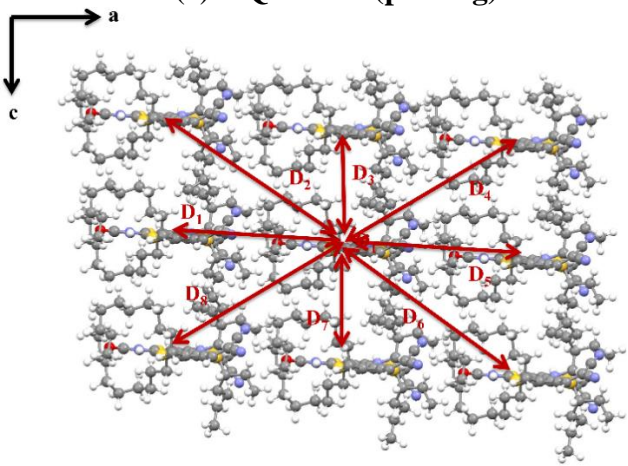
(e) **AQx-2-ct8 (packing)**

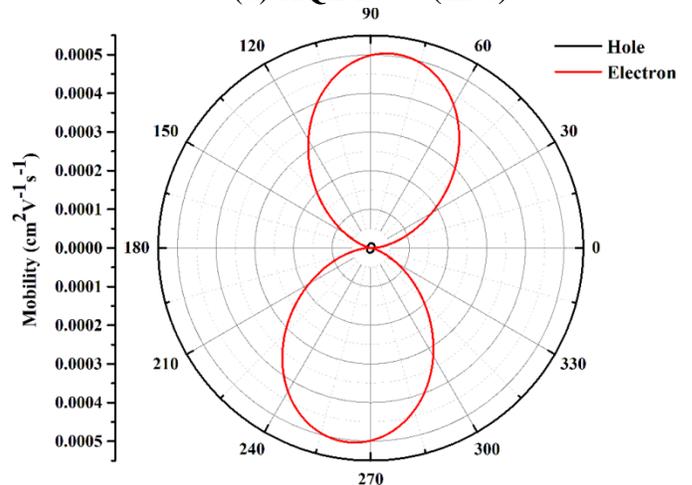
(f) **AQx-2-ct8 (mob)**

**Figure S12.** Simulated crystal structures showing different hopping channels and angle resolved anisotropic mobility of the studied NFAs (**AQx-2-ct6**, **AQx-2-ct7**, and **AQx-2-ct8**)

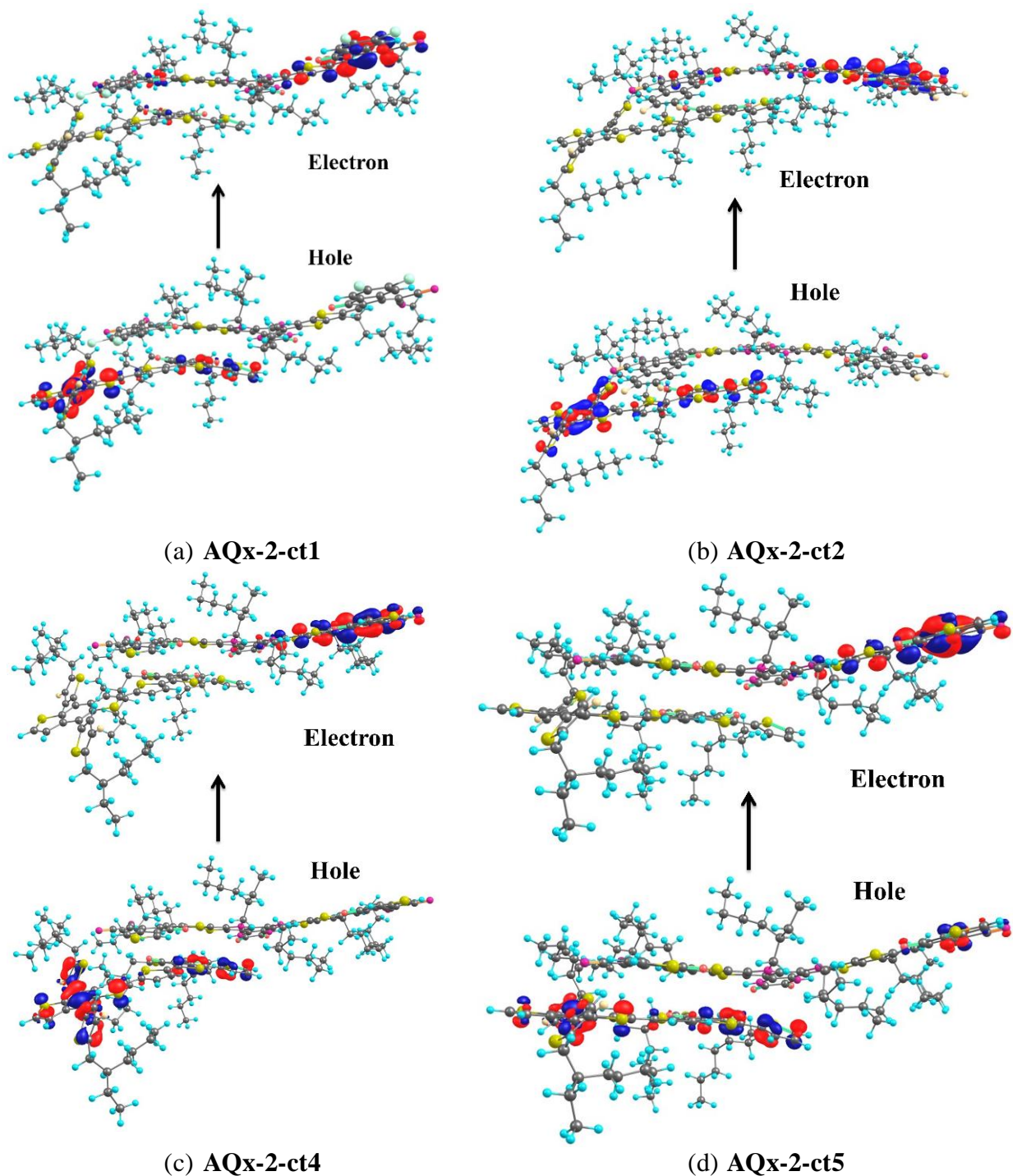

(a) **AQx-2-ct1**  (b) **AQx-2-ct2**

(c) **AQx-2-ct4**  (d) **AQx-2-ct5**

**Figure S13.** The NTO analysis of the CT excited states of the investigated PM6/NFA (**AQx-2-ct1**, **AQx-2-ct2**, **AQx-2-ct4**, **AQx-2-ct5**) blends. The blue and red colors represent the positive and negative isosurfaces, respectively

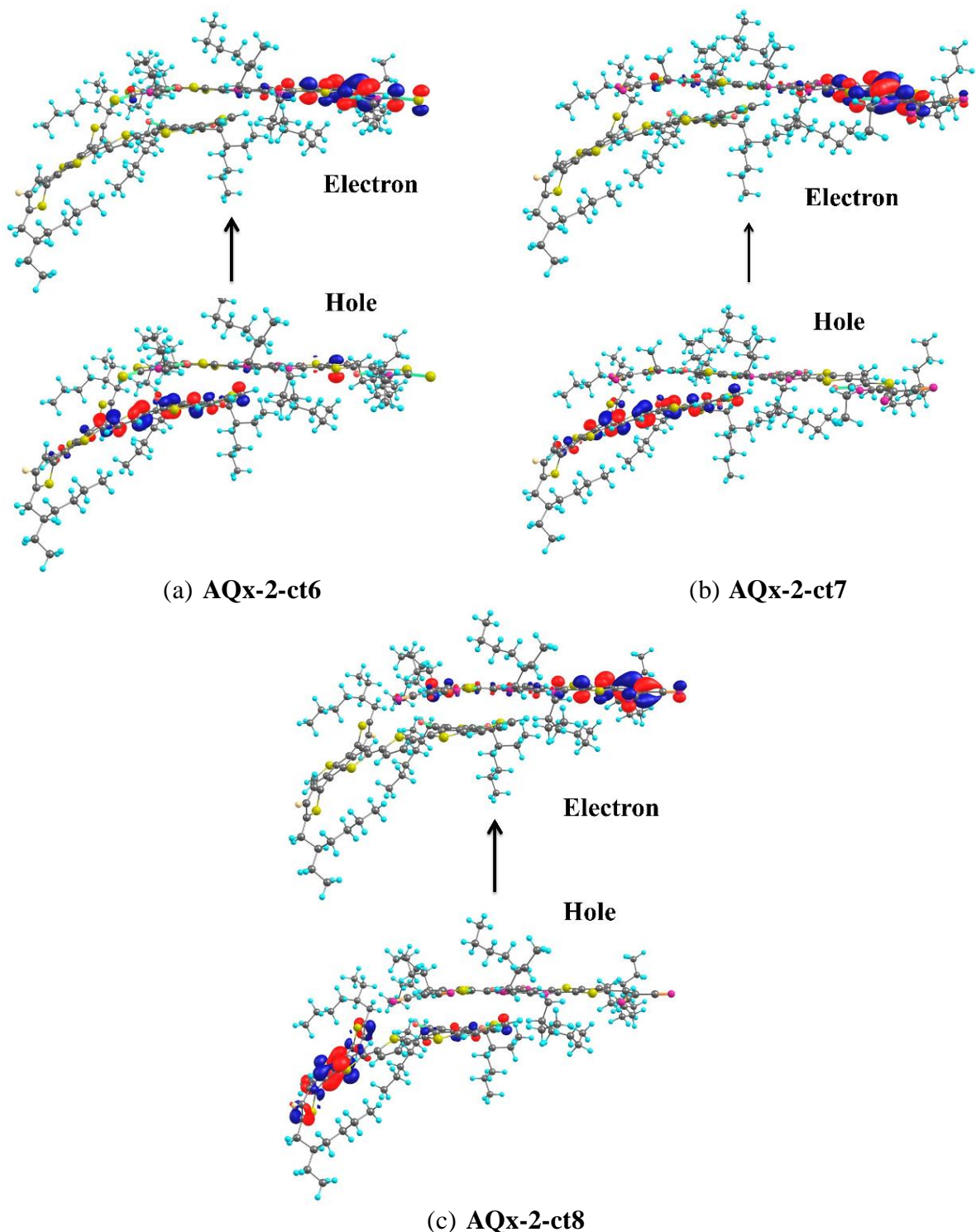

(a) **AQx-2-ct6**  (b) **AQx-2-ct7**

(c) **AQx-2-ct8**

**Figure S14.** The NTO analysis of the CT excited states of the investigated PM6/NFA (**AQx-2-ct6**, **AQx-2-ct7**, **AQx-2-ct8**) blends. The blue and red colors represent the positive and negative isosurfaces, respectively